\documentclass[final,times,twocolumn]{elsarticle}
\makeatletter
\def\ps@pprintTitle{%
 \let\@oddhead\@empty
 \let\@evenhead\@empty
 \def\@oddfoot{}%
 \let\@evenfoot\@oddfoot}
\makeatother
\usepackage{geometry}
\usepackage{lineno,hyperref,amsmath,amssymb,subcaption, nccmath, mathrsfs, mathtools, wasysym, float}
\usepackage{setspace}
\usepackage{siunitx}
\usepackage{cuted}

\DeclareSIUnit\wtpc{\text{wt\%}}

\modulolinenumbers[5]
\makeatletter
\newcommand*{\rom}[1]{\mathrm{\expandafter\@slowromancap\romannumeral #1@}}
\newcommand{\mr}[1]{\mathrm{#1}}
\newcommand{\h}[1]{\hat{#1}}
\newcommand{\td}[1]{\tilde{#1}}

\makeatother
\usepackage{layouts}
\usepackage{booktabs}
\usepackage{multirow}
\usepackage[table]{xcolor}

\begin{document}

\begin{frontmatter}

\title{Capillarity-driven thinning and breakup of weakly rate-thickening fluids} 


\author[addr1]{Jianyi Du}
\author[addr2]{Hiroko Ohtani}
\author[addr2]{Kevin Ellwood}
\author[addr1]{Gareth H. McKinley\corref{cor}}
\address[addr1]{Department of Mechanical Engineering, Massachusetts Institute of Technology, 77 Massachusetts Ave, Cambridge, MA 02139, USA}
\address[addr2]{Vehicle Manufacturing and Additive Manufacturing-Metals, Ford Motor Company, Dearborn, MI 48121, USA}
\cortext[cor]{Corresponding author (gareth@mit.edu)}

\begin{abstract}
A number of commercial fluids, including synthetic automotive oils, food and consumer products containing polymer additives exhibit weakly rate-thickening responses in the final stages of capillarity-driven thinning, where a large accumulated strain and high extensional strain rate alter the thinning dynamics of the slender liquid filament. Consequently, the capillarity-driven thinning dynamics typically feature two distinct regions at the early and late stages of the filament breakup process, each dominated by distinct mechanisms. These features have been incorporated in a simple Inelastic Rate-Thickening (IRT) model with linear and quadratic contributions to the constitutive stress-strain rate relationship, where the apparent extensional viscosity slowly thickens at high strain rates. We numerically compute the thinning dynamics of the IRT model assuming an axially-slender axisymmetric filament and no fluid inertia. The computational results motivate a new self-similar solution dominated by the second-order stress obtained through a similarity transformation. The new asymptotic solution leads to a self-similar filament shape that is more slender than the Newtonian counterpart and results in a quadratic thinning of the mid-point radius of the filament with time to breakup close to singularity. A new and distinct asymptotic geometric correction factor, $X\approx 0.5778$ is obtained, from which a more accurate true extensional viscosity can be recovered from an interpolated time-varying geometric correction factor based on the magnitudes of different stress components. Finally, we propose a statistics-based protocol to select the best-fit constitutive model using a parameter-free criterion, enabling us to quantify the extensional rheological behavior through capillarity-driven thinning dynamics more systematically on complex rate-thickening viscoelastic fluids. 
\end{abstract}


\end{frontmatter}


\section{Introduction}
Capillarity-driven thinning is prevalent in a number of rheological phenomena such as liquid filament breakup  \cite{mckinley2005visco}, jet impingement and spray atomization process \cite{keshavarz2015studying}. When an initially cylindrical liquid bridge is sufficiently stretched, it pinches off due to the action of surface tension into two separate drops. During the process a thin liquid filament forms between the two liquid reservoirs on both ends. The structural evolution of this transient liquid filament can be quantified by the radius of the filament $R(z,t)$, which is a function of axial position and time. From the measured evolution of $R(z,t)$, we can extract the transient extensional rheological properties of an unknown liquid sample \cite{mckinley2005visco}. A number of extensional rheometers based on this technique are available, such as the Capillary Breakup Extensional Rheometer (CaBER) \cite{bazilevsky1990liquid}, the Rayleigh-Ohnesorge Jetting Extensional Rheometer (ROJER) \cite{keshavarz2015studying} and dripping-on-substrate (ODES-DOS) devices \cite{dinic2015extensional}. These devices allow for the extensional rheological characterization of a wide range of materials, including Newtonian fluids \cite{papageorgiou1995breakup,mckinley2000extract}, dilute polymer solutions \cite{entov1997effect,clasen2006dilute}, emulsions \cite{niedzwiedz2010extensional} and particulate suspensions \cite{mcilroy2014modelling}. The extensional rheology of these materials is important in various applications such as paint spraying \cite{keshavarz2015studying}, oil recovery \cite{yamakov1997polymer}, droplet formation in microfluidics \cite{ahn2006dielectrophoretic}, turbulent drag reduction \cite{bhattacharjee1991drag}, and biomechanics \cite{juarez2011extensional}.

Capillarity-driven breakup techniques have provided a simple, yet effective method to characterize the transient extensional rheology of complex fluids described by a wide range of constitutive equations. However, to obtain accurate constitutive model parameters from the measured kinematics of the liquid filament, the temporal evolution of the net external axial force $F_\mr{a}$ exerted on a control volume element of the liquid thinning filament needs to be calculated \cite{mckinley2005visco}. 
Previous studies \cite{mckinley2000extract,clasen2006dilute} have used dimensional analysis to express this force as $F_\mr{a}=2\pi X\Gamma R_\mr{mid}(t)$, where $\Gamma$ is the surface tension of the fluid, and $R_\mr{mid}(t)=R(0,t)$ is the mid-plane radius of the liquid filament. The geometric correction factor $X$ accounts for the non-cylindrical slender profile of the thinning liquid filament that are determined by the interactions between capillarity and the different stress components arising from given constitutive equations. 
When the capillarity-driven thinning process is dominated by a single stress component in the constitutive equation, the geometric correction factor $X$ is generally constant. McKinley and Tripathi \cite{mckinley2000extract} showed that for a Newtonian fluid under the visco-capillary force balance, the magnitude of this correction factor has a non-trivial value of $X=0.7127$. This value can be rigorously obtained by assuming a self-similar shape of the liquid filament \cite{renardy1994some}. In the limit of elasto-capillary balance of the Oldroyd-B model, the liquid filament is close to a cylindrical shape due to the overwhelmingly large tensile force that arises from the axial elastic term. As a result, it was commonly assumed $X=1$ in the earlier studies \cite{entov1997effect, mckinley2005visco}. Admittedly, its accurate value is still in debate \cite{clasen2006beads}. 
For more complex constitutive models with multiple contributions to the total tensile stress, transitions in the dominant stress balance are to be expected broadly. Consequently, the geometric correction factor $X$ is not necessarily constant as the thinning liquid filament evolves, resulting in a more complex response of the external axial force $F_\mr{a}(t)$. If the material is only weakly viscoelastic, magnitudes of the different stress components remain comparable during the filament thinning process, then the capillarity-driven thinning response may be determined by a balance between capillarity and several different stress contributions to the total stress given by the constitutive model. 

Recently, two commercially-available synthetic automotive lubricants have been studied using a customized CaBER system \cite{duImprovedCapillaryBreakup2021,Du2022thesis}. Both fluids appear to be identically Newtonian in shear flow. However, they exhibit substantially distinct rate-thickening behaviors at large strain rates in a strong extensional flow. These two automotive lubricants have low-concentrations of particulate or polymer additives (less than \SI{5}{\text{wt\%}}) dispersed in relatively viscous solvents, and are representative of a number of industrial fluids with complex rheological behavior at large strain rates ($\SI{100}{\s^{-1}}\leq\dot{\epsilon}\leq\SI{1000}{\s^{-1}}$) \cite{gunsel1996friction}. 
The material response of the more-strongly elastic lubricant could be characterized by the familiar Oldroyd-B model, in which an elasto-capillary balance progressively replaces the initial visco-capillary balance in the axial direction and governs the dynamics of the filament thinning close to breakup. The other lubricant, however, shows more weakly rate-thickening behavior, and the extensional viscosity only slowly varies as the strain rate increases. To better describe the filament thinning dynamics of this weakly rate-thickening fluid, an Inelastic Rate-Thickening (IRT) model was proposed \cite{duImprovedCapillaryBreakup2021}, in which two constitutive parameters characterize the zero-shear-rate viscosity and the rate of extensional thickening, respectively.
However, a comprehensive understanding of the correct value of the geometric correction factor $X(t)$ that incorporates multiple stress contributions to the IRT model is not yet available. As a result, an accurate computation of the transient extensional viscosity at medium to large strain rates is so far unattainable. In the present paper,
a dimensionless ``1+1'' form of the coupled momentum and constitutive equations is solved numerically to understand the capillarity-driven thinning dynamics of the IRT model. The apparent extensional viscosity and the geometric correction factor $X(t)$ can be extracted from the computed evolution in the filament profiles. By inspecting the temporal evolution of $X$, we observe that as the filament thins and the strain rate (or the dimensionless Weissenberg number) increases, the magnitude of $X$ deviates from the viscocapillary solution for Newtonian fluids and asymptotically approaches a new constant. This new asymptotic solution originates from the growing second-order contribution to the stress in the IRT model close to filament breakup, and can be calculated analytically by assuming a new and more slender self-similar solution of the evolving liquid filament shape. 
To render a robust criterion for selecting the best-fit constitutive model from the measured evolution of the liquid filament profile with experimental noise, we implement the parameter-free Bayesian information criterion to evaluate different models based on both the mean square errors and the number of constitutive parameters. We incorporate this criterion to a more general testing protocol, which considers both the evolution of the geometric correction factor $X(t)$, as well as the choice of the best-fit model, to systematically extract the extensional rheological parameters from the measured filament thinning dynamics of a number of complex viscoelastic fluids.

\section{Model setup}
\label{sec:setup}
\begin{figure}[H]
\centering
\includegraphics[width=0.5\textwidth]{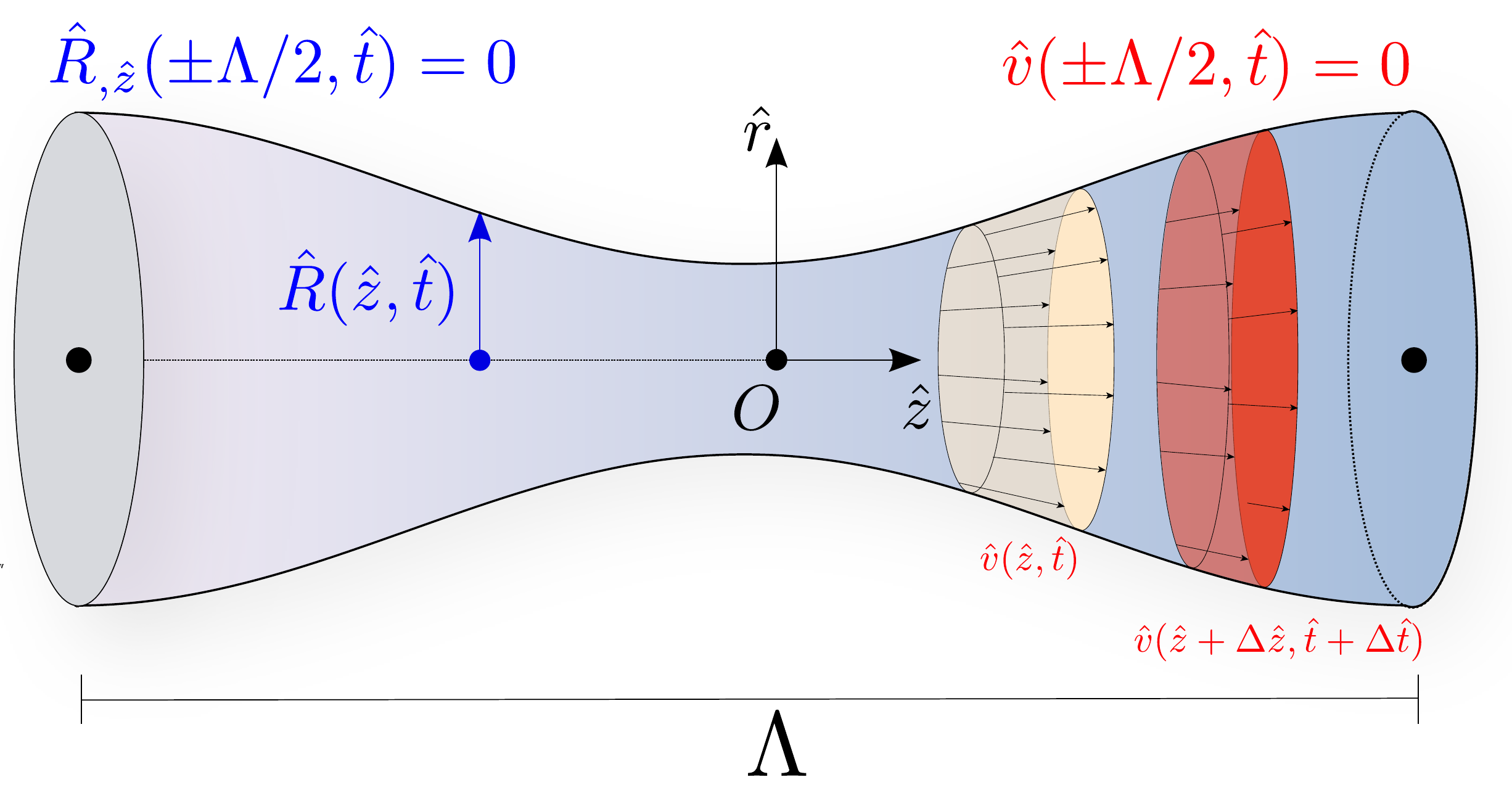}
\caption{Geometric setup for the numerical calculation of the capillarity-driven thinning dynamics on a capillary breakup rheometer. An axisymmetric slender piece of fluid filament with radius $\h{R}(\h{z},\h{t})$ and axial velocity $\h{v}(\h{z},\h{t})$ is confined within a distance of $\Lambda$. All quantities have been nondimensionalized using the disc radius $R_0$ (not shown) and the visco-capillary timescale of the fluid $t_\mr{vis}$. The boundary conditions are imposed as $\h{R}_{,\h{z}}(\pm\Lambda/2,\h{t})=0$ and $\h{v}(\pm\Lambda/2, \h{t})=0$.}
\label{fig:geometry}
\end{figure}

The filament thinning dynamics studied in this paper can be modeled as shown in Fig.~\ref{fig:geometry}. Here, a fluid sample with an axisymmetric shape around the axial $z$-axis is initiated with a length of $L$. On a capillary breakup extensional rheometer, this fluid sample is analogous to the liquid filament formed between two end reservoirs, and its filament radius $R=R(z,t)$ is a function of the axial position and time. In practice, each end reservoir is attached to a disc with a fixed radius $R_0$ (not shown in Fig.~\ref{fig:geometry}), which sets the maximum value of the initial mid-plane radius $R(0, 0)\sim R_0$. Therefore, $R_0$ acts as a characteristic length scale of the filament in the radial direction. 
In the filament, the local velocity in the axial direction $v(r,z,t)$ is assigned in a Eulerian framework and its magnitude varies both temporally and spatially (in both axial and radial directions). Without loss of generality, we use the disc radius $R_0$ and the visco-capillary timescale of the fluid $t_\mr{vis}$ to nondimensionalize all the quantities and operators, the latter of which is defined as
\begin{equation}
t_\mr{vis}\equiv\dfrac{\eta_0 R_0}{\Gamma},
\end{equation}
where the zero-shear viscosity $\eta_0$ is used to eliminate any strain or strain-rate dependence, and $\Gamma$ is the surface tension of the fluid. The following nondimensionalizing schemes are used throughout this study unless specified:
\begin{equation}
\begin{aligned}
~\h{r} &\equiv \dfrac{r}{R_0},
~\h{z}  \equiv \dfrac{z}{R_0},
~\h{t}  \equiv \dfrac{t}{t_\mr{vis}}, \\
~\h{R} &\equiv \dfrac{R}{R_0},
~\h{v}  \equiv \dfrac{v}{R_0/t_\mr{vis}},
~\Lambda  \equiv \dfrac{L}{R_0}, \\
~\h{\nabla} &\equiv R_0\nabla,
~ \partial_{\h{t}} \equiv t_\mr{vis}\partial_{t},
~ \partial_{\h{z}} \equiv R_0\partial_{z}.\\
\end{aligned}
\label{eqn:nondim}
\end{equation}
Here, variables on the left hand sides of each formula are dimensionless. The dimensionless length of the filament $\Lambda$ acts as an input aspect ratio. 
When a liquid profile is sufficiently slender, or $\Lambda\gg 1$, Eggers \cite{eggers1994drop} used the perturbation analysis to rigorously prove that $v(r,z,t)$ has a higher leading order regarding to the radial position $r$ than the axial position $z$. Therefore, we can approximate the axial velocity in the liquid filament to be radially uniform (as shown by the velocity profile in Fig.~\ref{fig:geometry}), or mathematically, the kinematics of the liquid filament can be quantified by $R=R(z,t)$ and $v=v(z,t)$. In addition, the boundary conditions are consistent with the setup of a capillary breakup extensional rheometer in dimensionless forms as
\begin{subequations}
\begin{align}
\h{R}_{,\h{z}}(\pm \Lambda/2, \h{t})&=0,\\
\h{v}(\pm \Lambda/2,\h{t})&=0,
\end{align}
\label{eqn:bdry}%
\end{subequations}
where the notation of ``$f_{,x}$'' indicates the partial derivative of $f$ with respect to $x$. These boundary conditions are clearly marked in Fig.~\ref{fig:geometry}.

To initiate the capillarity-driven thinning, we impose an appropriate initial condition for the shape and velocity of the liquid filament following the criterion proposed by
Slobozhanin and Perales \cite{slobozhanin1993stability} as well as Papageorgiou \cite{papageorgiou1995breakup} as
\begin{subequations}
\begin{align}
\h{R}(\h{z},0)&=0.5-0.1\cos{\left(\dfrac{2\pi \h{z}}{\Lambda}\right)},\\
\h{v}(\h{z},0)&=0.
\end{align}
\label{eqn:init}%
\end{subequations}
When the filament is slender, Eggers and Dupont \cite{eggers1994drop} rewrote the momentum equation in the absence of radial dependence as 
\begin{equation}
\left(\h{R}^{2} \h{v}\right)_{,\h{t}}+\left(\h{R}^{2} \h{v}^{2}\right)_{,\h{z}}=\left[\h{R}^{2}\left(\h{K}+3\h{v}_{,\h{z}}+\Delta \h{\sigma}^{nN}+\frac{\dot{\h{R}}^{2}}{2}\right)\right]_{,\h{z}}-\h{R}^{2} \cdot\mr{Bo},
\label{eqn:euler_force}
\end{equation}
where the subscripts $\h{t}$ and $\h{z}$ represent the partial derivatives with respect to time and axial position, respectively. The dimensionless Bond number is defined as $\mr{Bo}\equiv (\rho g R_0^2)/\Gamma$, which quantifies the gravitational-capillary effect parameterizing the fluid density $\rho$, gravitational acceleration $g$, the characteristic radial length $R_0$ and the surface tension $\Gamma$. The term with $3\h{v}_{,\h{z}}$ represents the dimensionless viscous stress term arising from the Newtonian contribution, followed by the first normal stress difference of the non-Newtonian terms in the constitutive equation, $\Delta \h{\sigma}^{nN}=\h{\sigma}_{zz}^{nN}-\h{\sigma}_{rr}^{nN}$. The fluid inertia is represented by the term $\dot{\h{R}}^2/2\equiv(\partial_{\h{t}} \h{R})^2/2$. The dimensionless capillary pressure $\h{K}$ can be expressed using Young-Laplace equation as
\begin{equation}
\h{K}=\frac{1}{\h{R}\left(1+\h{R}_{,\h{z}}^2\right)^{1/2}}+\frac{\h{R}_{,\h{z}\h{z}}}{\left(1+\h{R}_{,\h{z}}^2\right)^{3/2}}.
\label{eqn:yl_eqn}
\end{equation}
Finally, the continuity equation can be written as
\begin{equation}
\left(\h{R}^2\right)_{,\h{t}} + \left(\h{v}\h{R}^2\right)_{,\h{z}}=0.
\label{eqn:free_surface}
\end{equation}

In the capillary breakup techniques, the minimal filament radius is a measure of the evolution of capillarity-driven filament thinning, from which a number of extensional rheological properties can be extracted \cite{mckinley2005visco}. When the gravitational effects are negligible ($\mr{Bo}\ll 1$), this quantity is identical to the mid-plane radius, \textit{i.e.}, $\h{R}_\mr{mid}(\h{t})\equiv \h{R}(0,\h{t})$. Although Eq.~\ref{eqn:euler_force} to \ref{eqn:free_surface} provide a closed-form expression of the liquid filament shape, we need to numerically solve the full partial differential equations in order to obtain $\h{R}_\mr{mid}(\h{t})$. To simplify the calculation, a stress balance equation at the mid-plane is commonly applied derived from the control volume analysis as \cite{mckinley2000extract}
\begin{equation}
\mr{Tr}\cdot \mr{Wi}\equiv 3\mr{Wi}+\Delta \h{\sigma}^{nN}(0,\h{t})=\frac{\h{F}_a(\h{t})}{\pi \h{R}^{2}_\mr{mid}} - \h{K}(0,\h{t}) . 
\label{eqn:lag_force}
\end{equation}
On the left-hand side of this equation, we subsume the resistance of filament thinning from the liquid to a dimensionless extensional viscosity, or the Trouton ratio $\mr{Tr}\equiv\eta_\mr{e}/\eta_0$, where $\eta_\mr{e}$ and $\eta_0$ are the extensional and zero-shear viscosities, respectively. The dimensionless strain rate at $z=0$, known as the Weissenberg number is defined as $\mr{Wi}\equiv \h{v}_{,\h{z}}(0,\h{t})=-2(\h{R}_\mr{mid})_{,\h{t}}/\h{R}_\mr{mid}$. On the right-hand side, the resistance is balanced by the capillary pressure at mid-plane $\h{K}(0,\h{t})$ and an external axial force (dimensionless) $\h{F}_\mr{a}(t)$ exerted by the fluid reservoirs on both ends. From dimensional analysis, this external force can be expressed as $F_\mr{a}(t)=2\pi X \Gamma R(0, t)$, where $X$ is a geometric correction factor to account for the non-cylindrical shape of the liquid filament. We nondimensionalize this force term by $\h{F}_\mr{a}\equiv F_\mr{a}/(\Gamma R_0)=2\pi X \h{R}_\mr{mid}$. To extract accurate extensional rheological properties from the evolution of dimensionless mid-plane radius $\h{R}_\mr{mid}(\h{t})$, we need to determine the magnitude of $X$ for the applied constitutive model. For a Newtonian fluid ($\Delta \sigma^{nN}=0$), McKinley and Tripathi \cite{mckinley2000extract} showed that 
\begin{equation}
\h{R}_\mr{mid}(\h{t})=\h{R}_\mr{mid}(0)-\frac{2X_\mr{N}-1}{6}\h{t},
\label{eqn:newtonian}
\end{equation}
where the geometric correction factor $X_\mr{N}\approx 0.7127$. This value can also be analytically obtained by substituting a self-similar ansatz of the liquid filament shape \cite{renardy1994some}. For a viscoelastic fluid governed by the Oldroyd-B model under an elasto-capillary balance, Entov and Hinch \cite{entov1997effect} 
assumed the normal stress in $zz$ direction, $\sigma_{zz}$ to vanish, thus the value of $X$ to be unity. Clasen et al. \cite{clasen2006beads} corrected this value by justifying a non-trivial value of $\sigma_{zz}$ arising from the bulk fluid, which leads to a larger value of $X=3/2$. The resulting evolution of $\h{R}_\mr{mid}(\mr{t})$ deviates from the study by Entov and Hinch by a factor of $2^{-1/3}$.

In this study, we primarily focus on weakly rate-thickening fluids. This type of fluids appears to be Newtonian fluids in shear flow, but can exhibit rate-thickening behavior at large strain rates in a strong extensional flow. Du et al. \cite{duImprovedCapillaryBreakup2021} have measured the extensional rheology of two synthetic automotive lubricants on a customized CaBER system. An Inelastic Rate-Thickening (IRT) model was proposed to successfully describe both the shear and extensional rheology of the weakly rate-thickening lubricants. The two-parameter IRT model consists of two stress components and can be expressed as

\begin{subequations}
\begin{align}
\boldsymbol{\sigma}=\eta(\rom{2}_{\boldsymbol{\dot{\gamma}}},\rom{3}_{\boldsymbol{\dot{\gamma}}})\boldsymbol{\dot{\gamma}},\\
\eta(\rom{2}_{\boldsymbol{\dot{\gamma}}},\rom{3}_{\boldsymbol{\dot{\gamma}}})=\eta_0+k_2\dot{\epsilon},
\end{align}
\label{eqn:sof}%
\end{subequations}
where $\boldsymbol{\sigma}$ and $\boldsymbol{\dot{\gamma}}$ are the stress and strain-rate tensors. The rate-dependent viscosity $\eta$ is parameterized by the zero-rate viscosity $\eta_0$ and the rate of thickening $k_2$, and can be written as a function of the second and third invariants of $\boldsymbol{\dot{\gamma}}$, which are defined as $\rom{2}_{\boldsymbol{\dot{\gamma}}}=\mr{tr }(\boldsymbol{\dot{\gamma}}\cdot \boldsymbol{\dot{\gamma}})$ and $\rom{3}_{\boldsymbol{\dot{\gamma}}}=\mr{tr }(\boldsymbol{\dot{\gamma}\cdot\dot{\gamma}\cdot \dot{\gamma}})$ \cite{bird1987dynamics}. From these two invariants, a characteristic extensional rate is defined as $\dot{\epsilon}=\rom{3}_{\boldsymbol{\dot{\gamma}}}/\rom{2}_{\boldsymbol{\dot{\gamma}}}$ \cite{debbaut1988extensional}.

To substitute Eq.~\ref{eqn:sof} into the dimensionless forms of the momentum equations, we nondimensionalize the constitutive equation according to Eq.~\ref{eqn:nondim}, which results in a dimensionless elasto-capillary number in the expression of the rate-dependent viscosity as 
\begin{equation}
\mr{Ec}_0\equiv\frac{k_2}{\eta_0 t_\mr{vis}}=\dfrac{k_2\Gamma}{\eta_0^2 R_0},
\label{eqn:de}
\end{equation}
and $\eta(\rom{2}_{\boldsymbol{\dot{\gamma}}},\rom{3}_{\boldsymbol{\dot{\gamma}}})/\eta_0=1+\mr{Ec}_0\cdot\mr{Wi}$. This model presents the simplest form of a constitutive equation retaining a rate-thickening feature in an extensional flow. Du et al. \cite{duImprovedCapillaryBreakup2021} has rigorously proved the mathematical equivalence of the IRT model and the Oldroyd-B model in the limit of $\mr{Wi}\ll 1$, as well as the equivalence to the second-order fluid (SOF) model in a slow and slowly varying extensional flow. 

\section{Self-similar solutions}
\subsection{Numerical calculation}
\label{sec:self-similar}
Due to the second-order stress component dominating the capillarity-driven thinning dynamics in the IRT model close to filament breakup, we anticipate the overall geometric correction factor $X$ to deviate from the solution of $X_\mr{N}=0.7127$ for a Newtonian fluid as the strain rate increases. To fully understand the temporal evolution of the geometric correction factor during the filament thinning process, we substitute the constitutive equation (Eq.~\ref{eqn:sof}) into Eq.~\ref{eqn:euler_force} to~\ref{eqn:free_surface}, and numerically calculate the temporal evolution of the filament thinning profiles. This calculation is performed following the numerical procedures of Eggers and Dupont \cite{eggers1994drop}, in which we discretize the filament radius $\h{R}^{n,i}$ and axial velocity $\h{v}^{n,i+1/2}$ at $n$-th time step on grid nodes $i$ and vertices $i+1/2$ between node $i$ and $i+1$, respectively. Therefore, the Newtonian and non-Newtonian stress components in Eq.~\ref{eqn:euler_force} for the IRT model are explicitly expressed in a finite-difference form as
\begin{subequations}
\begin{align}
\left[\h{R}^2\left(3\h{v}_{,\h{z}}\right)\right]_{,\h{z}}^{n,i}&\equiv \frac{3(\h{R}^{n,i+1})^2\h{v}_{,\h{z}}^{n,i+1}-3(\h{R}^{n,i})^2\h{v}_{,\h{z}}^{n,i}}{\h{z}^{i+1}-\h{z}^{i-1}},\\
\left[\h{R}^2\left(\Delta \h{\sigma}^{nN}\right)\right]_{,\h{z}}^{n,i}&\equiv \frac{3\mr{Ec}_0\left[(\h{R}^{n,i+1})^2\h{v}_{,\h{z}\h{z}}^{n,i+1}-(\h{R}^{n,i})^2\h{v}_{,\h{z}\h{z}}^{n,i}\right]}{\h{z}^{i+1}-\h{z}^{i-1}}.
\end{align}
\label{eqn:nvt}%
\end{subequations}
Here, the first- and second-order partial derivatives of the axial velocity with regard to the axial position are defined on the grid nodes, and they are calculated according to
\begin{subequations}
\begin{align}
\h{v}_{,\h{z}}^{n,i}&\equiv \dfrac{\h{v}^{n,i+1/2}-\h{v}^{n,i-1/2}}{\h{z}^{i+1/2}-\h{z}^{i-1/2}},\\
\h{v}_{,\h{z}\h{z}}^{n,i}&\equiv \dfrac{\h{v}_{,\h{z}}^{n,i+1}-\h{v}_{,\h{z}}^{n,i-1}}{\h{z}^{i+1}-\h{z}^{i-1}}.
\end{align}
\label{eqn:vz_vzz}%
\end{subequations}
We can thus rewrite Eq.~\ref{eqn:euler_force} and \ref{eqn:free_surface} in finite-difference forms, and rearrange the time derivative terms in both equations to the left-hand side. A semi-implicit scheme is applied to evaluate the right-hand side at a ``mid-step'' radius $\h{R}^{n,i}_\mr{RHS}$ with an axial velocity $\h{v}^{n,i}_\mr{RHS}$ defined as
\begin{subequations}
\begin{align}
\h{R}^{n,i}_\mr{RHS}&=\h{R}^{n,i}+\Theta (\h{R}^{n+1,i}-\h{R}^{n,i}),\\
\h{v}^{n,i}_\mr{RHS}&=\h{v}^{n,i}+\Theta (\h{v}^{n+1,i}-\h{v}^{n,i}).
\end{align}
\label{eqn:step}%
\end{subequations}
Here, $0\leq\Theta\leq 1$ is a stride parameter. By letting $\Theta=0$ or $1$, we recover the fully explicit or implicit finite-difference forms. In our numerical calculation, we set $\Theta=0.55$. This number is consistent with Eggers and Dupont \cite{eggers1994drop} to produce smoother discrete solutions while keeping the leading-order truncation errors sufficiently small. The number of grid nodes is set to $N=128$, and the time step is $\Delta \h{t}=0.01$. 

\begin{figure*}[!]
    \centering
    \includegraphics[width=1\textwidth]{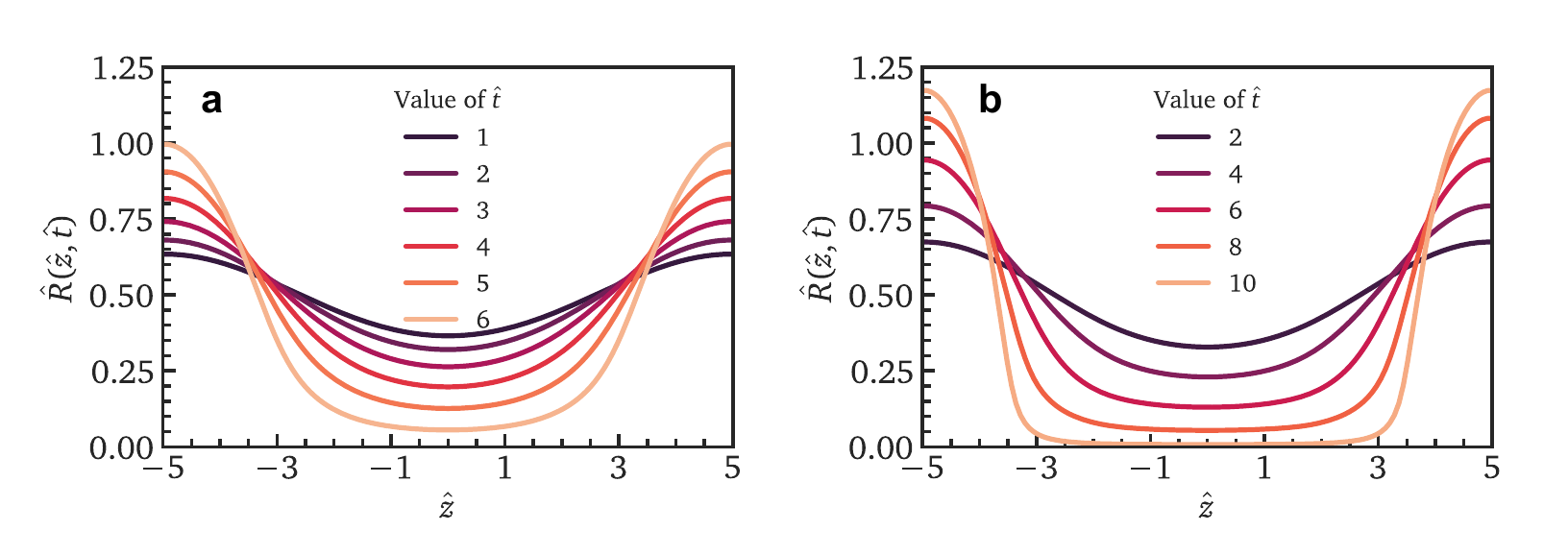}
    \caption{The filament profiles $\h{R}(\h{z},\h{t})$ at different time shown in the legends for (a) $\mr{Ec}_0=0$, and (b) $\mr{Ec}_0=0.5$. The number of nodes in the numerical calculation is $N=128$ and the time step is $\Delta \h{t}=0.01$. The liquid filament has the length of $\Lambda=10$. Filament profiles with a high elasto-capillary number in (b) are more slender in shape due to the additional second-order stress.}
    \label{fig:h_profile}
\end{figure*}

In the previous study \cite{duImprovedCapillaryBreakup2021}, the experimental results on weakly rate-thickening lubricants showed that the magnitudes of both linear and second-order stresses remain comparable during the filament thinning. To keep this physical fidelity, we let $\mr{Ec}_0\lesssim 1$ in our numerical calculation. The filament length is set as $\Lambda=10$, and the initial condition is specified in Eq.~\ref{eqn:init}. Fig.~\ref{fig:h_profile}(a) and (b) show the liquid filament profiles $\h{R}(\h{z},\h{t})$ with $\mr{Ec}_0=0$ (Newtonian fluid model) and $\mr{Ec}_0=0.5$ at different time. We compare the filament profiles of the two elasto-capillary numbers with similar mid-plane radii $\h{R}_\mr{mid}(\h{t})$ 
and find that the filament profiles for $\mr{Ec}_0=0.5$ are more slender in shape than those for $\mr{Ec}_0=0$. Such difference originates from the additional second-order stress in the constitutive equation, which increases quadratically in magnitude with the strain rate and overtakes the Newtonian stress in finite time, progressively dominating the filament thinning dynamics. 

\begin{figure*}[t!]
    \centering
    \includegraphics{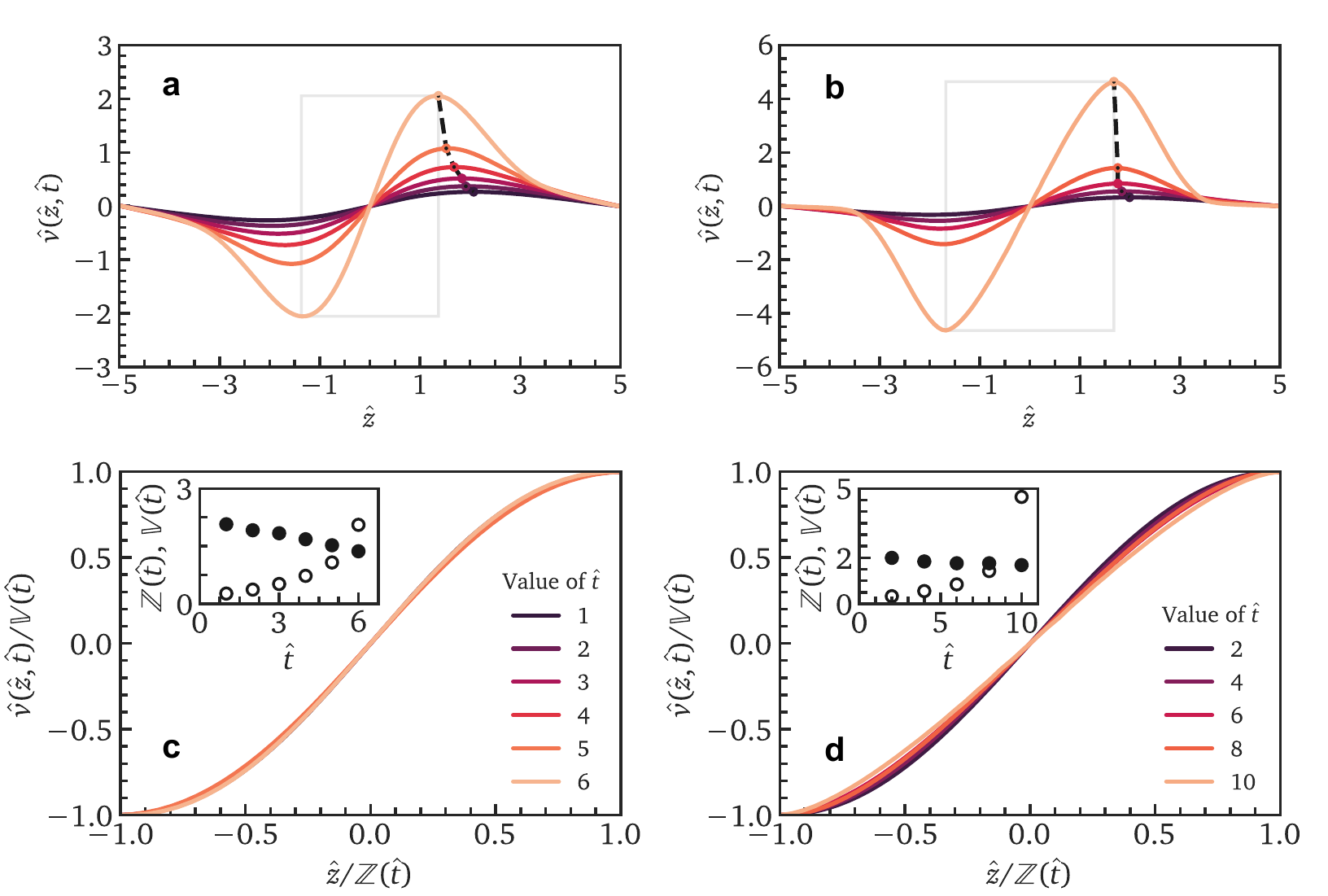}

    \caption{The axial velocity profiles $\h{v}(\h{z},\h{t})$ at different times $\h{t}$ for (a) $\mr{Ec}_0=0$, and (b) $\mr{Ec}_0=0.5$. In (c) and (d), only $\h{v}(\h{z},\h{t})$ in the dashed boxes from (a) and (b) is shown, where both the abscissa and ordinate are normalized with two shift factors $\mathbb{Z}(\h{t})$ and $\mathbb{V}(\h{t})$, respectively. The two shifting factors are determined from the position where the strain rate is zero, or mathematically $\h{v}_{,\h{z}}(\h{z}, \h{t})=0$. Inset: temporal evolution of $\mathbb{Z}(\h{t})$ (filled symbols) and $\mathbb{V}(\h{t})$ (hollow symbols). In (c), the capillary thinning dynamics of a Newtonian fluid ($\mr{Ec}_0=0$) is self-similar, which results in a master curve for the reduced velocity profile at different time. In (d), when the elasto-capillary number is nonzero, the reduced velocity profiles at different time do not overlap due to the higher-order stress breaking the self-similar nature originally governed by the Newtonian stress.}
    \label{fig:vel}
\end{figure*}

We further plot the corresponding velocity profiles $\h{v}(\h{z},\h{t})$ in Figure~\ref{fig:vel}(a) and (b) for the two fluids with different elasto-capillary numbers. Two shifting factors $\mathbb{Z}(\h{t})$ and $\mathbb{V}(\h{t})$ are imposed to the abscissa and ordinate, such that the zero-rate position where $\h{v}_{,\h{z}}(\h{z},\h{t})=0$ is reduced to $[\h{z}/\mathbb{Z}(\h{t}), \h{v}/\mathbb{V}(\h{t})]=(\pm 1, \pm 1)$. The resulting velocity profiles after scaling are plotted in Fig.~\ref{fig:vel}(c) and (d) in the window of $-1\leq \h{z}/\mathbb{Z}(\h{t})\leq 1$ and $-1\leq \h{v}/\mathbb{V}(\h{t})\leq 1$. In Fig.~\ref{fig:vel}(c) ($\mr{Ec}_0=0$), the reduced velocity profiles at different time overlap with each other, which indicates that the capillary thinning for a Newtonian fluid is self-similar. In Fig.~\ref{fig:vel}(d) ($\mr{Ec}_0=0.5$), however, the reduced velocity profile progressively reshapes with time. Compared with Fig.~\ref{fig:h_profile}, this evolution of the reduced velocity profile shows that the second-order stress in the IRT model results in a new pattern of the capillarity-driven thinning dynamics, which breaks the self-similar solution originally set up by the Newtonian stress. 

\begin{figure*}[t!]
    \centering
    \includegraphics{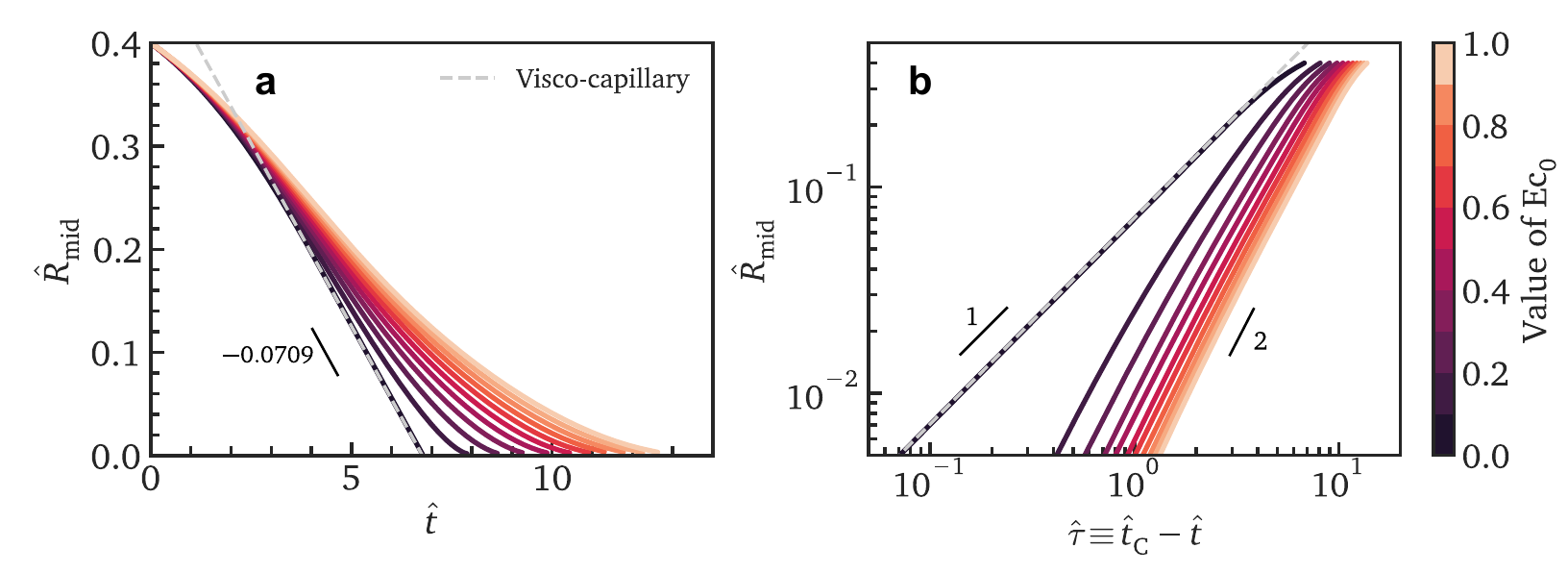}
    \caption{The mid-plane radius of the filament $\h{R}_\mr{mid}(\h{t})=\h{R}(0,\h{t})$ against dimensionless time $\h{t}$ and time to breakup $\h{\tau}$ for $0\leq \mr{Ec}_0\leq 1$. In (a), the mid-plane radius $\h{R}_\mr{mid}$ is plotted against time $\h{t}$. For $\mr{Ec}_0=0$, $\h{R}_\mr{mid}$ decays linearly with time close to breakup. The slope in the linear decay region is consistent with the previous solution of Newtonian fluids in Eq.~\ref{eqn:newtonian} as \num{-0.0709}. As $\mr{Ec}_0$ becomes positive, the capillarity-driven breakup is retarded due to the additional resistance from the second-order stress, and the evolution of the mid-plane radius deviates from a linear decay close to breakup. In (b), $\h{R}_\mr{mid}$ is plotted against the time difference $\h{\tau}\equiv\h{t}_\mr{C}-\h{t}$, where $\h{t}_\mr{C}$ is the breakup time determined by extrapolating $\h{R}_{mid}$ to vanish. For a Newtonian fluid ($\mr{Ec}_0=0$), a linear decay of $\h{R}_\mr{mid}(\h{\tau})$ is manifested as $\h{\tau}\rightarrow 0$. Whereas for $\mr{Ec}_0>0$, a quadratic trend of the mid-plane radius is observed.}
\label{fig:mid_plane_h}
\end{figure*}

To quantify the capillary thinning dynamics governed by the Newtonian and second-order stresses in the IRT model, we further extract the evolution of mid-plane radius $\h{R}_\mr{mid}(\h{t})$ from Fig.~\ref{fig:h_profile} with a range of elasto-capillary numbers $0\leq\mr{Ec}_0\leq 1$. In Fig.~\ref{fig:mid_plane_h}(a) and (b), the mid-plane radius $\h{R}_\mr{mid}(\h{t})$ is plotted against $\h{t}$ and $\h{\tau}$, respectively. Here, $\h{\tau}\equiv\h{t}_\mr{C}-\h{t}$ is the difference between time $\h{t}$ and the filament breakup time $\h{t}_\mr{C}$. Because $\h{t}=\h{t}_\mr{C}$ is a singularity of the solution, we calculate this filament breakup time by extrapolating the mid-plane radius to $\h{R}_\mr{mid}=0$. In Fig.~\ref{fig:mid_plane_h}(a), the mid-plane radius for a Newtonian fluid (dark line, $\mr{Ec}_0=0$) decays linearly close to breakup. This linear trend is consistent with the asymptotic solution for a Newtonian fluid from Eq.~\ref{eqn:newtonian} (dashed line) with a slope of $-(2X_\mr{N}-1)/6=-0.0709$. When $\mr{Ec}_0>0$, however, the additional second-order stress in the liquid filament leads to an apparent breakup retardation, as shown in Fig.~\ref{fig:mid_plane_h}(a), and the filament breakup time $\h{t}_\mr{C}$ becomes larger as $\mr{Ec}_0$ increases. To visualize the asymptotic evolution of the mid-plane radius governed by the second-order stress, we replot $\h{R}_\mr{mid}$ against $\h{\tau}$ on a logarithmic scale, as shown in \ref{fig:mid_plane_h}(b). Here, when $\mr{Ec}_0>0$, the mid-plane radius decays quadratically with $\h{\tau}$ close to breakup. This quadratic trend suggests a new asymptotic solution when the balance between capillarity and the second-order stress in the IRT model dominates the filament thinning dynamics.

\begin{figure}[!h]
    \centering
    \includegraphics[width=0.5\textwidth]{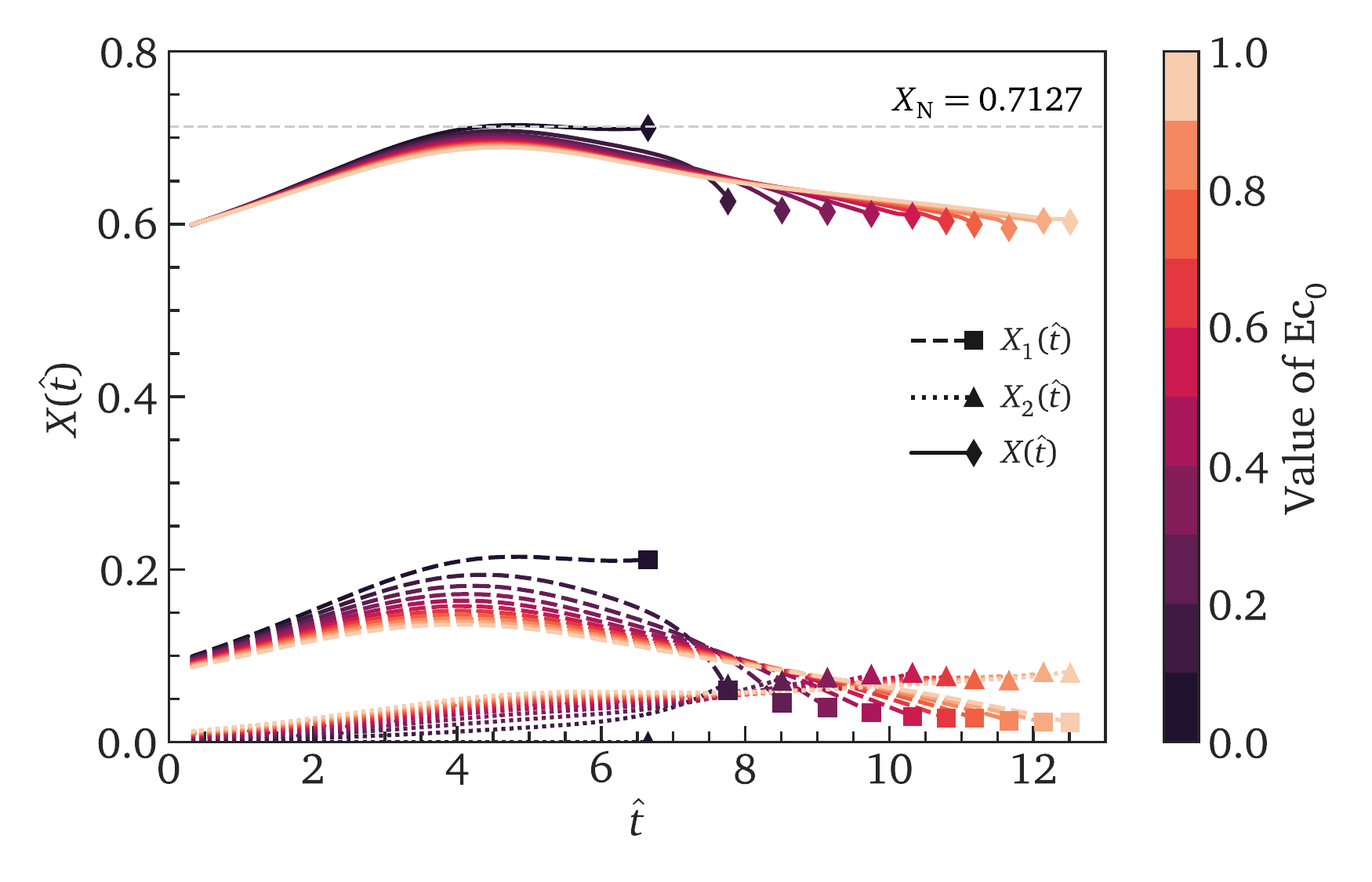}
    \caption{Contributions of the linear and second-order stress components in the IRT model to the geometric correction factor $X(\h{t})$, denoted as $X_1(\h{t})$ and $X_2(\h{t})$, as well as the overall geometric correction factor plotted against dimensionless time $\h{t}$ at different elasto-capillary numbers. The contribution of the capillary force $X_\mr{cap}$ is a constant of $0.5$ (not plotted). For a Newtonian fluid ($\mr{Ec}_0=0$, the darkest lines), the overall geometric correction factor reaches a constant which agrees with the previous solution under a visco-capillary balance as $X_\mr{N}\approx 0.7127$ close to breakup (gray dashed line) \cite{mckinley2000extract}. When $\mr{Ec}_0>0$, the contribution of the linear stress component $X_1(\h{t})$ (squares) reaches a local maximum at $\h{t}\approx 5$, and then progressively vanishes. In contrast, the contribution of the second-order stress component $X_2(\h{t})$ (triangles) steadily increases and surpasses $X_1(\h{t})$ in finite time. The overall geometric correction factor $X$ (diamonds) asymptotically approaches a new constant $X_\mr{RT}$ smaller than $X_\mr{N}$ close to breakup.}
    \label{fig:diffDe_force}
    \end{figure}

From the numerical calculations in Fig.~\ref{fig:h_profile}, we can extract the geometric correction factor $X$. In Eq.~\ref{eqn:lag_force}, for the IRT model, the external axial force $\h{F}_{a}(\h{t})=2\pi X \h{R}_\mr{mid}$ summates three force components of the stress balance equation, which can be expressed as
\begin{equation}
\h{F}_\mr{a}(\h{t})=\h{F}_\mr{cap}(\h{t})+
                     \h{F}_1(\h{t})+\h{F}_2(\h{t}),
\end{equation}
where $\h{F}_\mr{cap}$ is the capillary force arising from Eq.~\ref{eqn:yl_eqn}, and $\h{F}_1$ and $\h{F}_2$ are the linear (Newtonian) and second-order stress components in the constitutive equation. The capillary force can be expressed as $\h{F}_\mr{cap}(\h{t})=\pi \h{R}_\mr{mid}$ at the mid-plane $\h{z}=0$. The contribution of the axial curvature from Eq.~\ref{eqn:yl_eqn} to the total force is not explicitly expressed in $\h{F}_\mr{cap}$ but subsumed into $X$ \cite{papageorgiou1995breakup}. The linear and second-order components can be expressed in a dimensionless form according to Eq.~\ref{eqn:sof} as
\begin{subequations}
\begin{align}
\h{F}_1(\h{t})&=3\pi \mr{Wi}\h{R}_\mr{mid}^2, \\
\h{F}_2(\h{t})&=3\pi \mr{Ec}_0\mr{Wi}^2\h{R}_\mr{mid}^2.
\end{align}
\end{subequations}
In Fig.~\ref{fig:diffDe_force}, we plot the contributions of each force component to the overall geometric correction factor $X$ as $X_i\equiv \h{F}_i/(2\pi \h{R}_\mr{mid})$, where $i=\{1,2,\text{``cap''}\}$. By this definition, the contribution from the capillary force $X_\mr{cap}$ remains constant of $0.5$ throughout the filament thinning process and is thus not presented. For the linear (Newtonian) stress component, the increase of $X_1(\h{t})$ (squares) from $\h{t}=0$ to $\h{t}\approx 5$ is primarily due to the filament acceleration from the initial condition and is thus a numerical artifact \cite{papageorgiou1995breakup}. Beyond a local maximum at $\h{t}\approx 5$, the evolution of $X_1(\h{t})$ is determined by the magnitude of the elasto-capillary number. When $\mr{Ec}_0=0$ (Newtonian fluid model), $X_1(\h{t})$ remains a constant and $X_2(\h{t})$. The overall geometric correction factor $X$ (diamonds) is reduced to the solution under a visco-capillary balance as $X_\mr{N}\approx 0.7127$ until filament breakup at $\h{t}_\mr{C}\approx 6.6$ \cite{mckinley2000extract}. When $\mr{Ec}_0>0$, however, the contribution of the linear stress $X_1(\h{t})$ progressively vanishes from the local maximum. The magnitude of the local maximum also decreases as $\mr{Ec}_0$ increases. In contrast, the contribution from the second-order stress $X_2(\h{t})$ (triangles) steadily increases and surpasses $X_1(\h{t})$ close to the filament breakup and dominates the remaining filament thinning process. As shown in Fig.~\ref{fig:diffDe_force}, the overall geometric correction factor $X(\h{t})$ at different positive elasto-capillary numbers (diamonds) consistently converge to another constant smaller than $X_\mr{N}$, denoted as $X_\mr{RT}$ (corresponding to the ``rate-thickening'' term). Such convergence reveals that the capillarity-driven thinning dynamics governed by the second-order stress follows a self-similar relation with a different set of parameters from the Newtonian fluid. 

From Fig.~\ref{fig:diffDe_force} and the quadratic evolution of the mid-plane radius $\h{R}_\mr{mid}(t)$ close to breakup in Fig.~\ref{fig:mid_plane_h}, the capillarity-driven thinning dynamics of a weakly rate-thickening fluid characterized by the IRT model are more complex than a Newtonian fluid due to the interplay of two distinct self-similar laws arising from the first- and second-order stresses, respectively. By substituting the unknown constant $X_\mr{RT}$ into Eq.~\ref{eqn:lag_force}, we can express the asymptotic evolution of the mid-plane radius $\h{R}_\mr{mid}$ for $\mr{Ec}_0>0$ close to breakup as
\begin{equation}
\h{R}_\mr{mid}(\h{\tau};\mr{Ec}_0)\rightarrow	\frac{2X_\mr{RT}-1}{48\mr{Ec}_0}\h{\tau}^2.
\label{eqn:h}
\end{equation}
This quadratic trend is consistent with the results from Fig.~\ref{fig:mid_plane_h} from the numerical calculations. A similar expression has been previously obtained by McKinley \cite{mckinley2005visco}, in which $X_\mr{RT}$ is assumed to be unity based on the assumption of a cylindrical filament profile. In this study, we obtain a non-trivial value of $X_\mr{RT}$ to evaluate the evolution of the mid-plane radius and the extensional viscosity more accurately.
\subsection{Analytical solution}
\label{sec:model}
Inspired by the numerical results from Sec.~\ref{sec:self-similar}, we intentionally separate the first- and second-order stress components in the IRT model to probe the individual capillarity-driven thinning behavior from each stress component. Renardy \cite{renardy1994some} first obtained the solution of a Newtonian fluid ($\mr{Ec}_0=0$) under a visco-capillary balance. The evolution of the mid-plane radius and axial velocity can be expressed according to a self-similar ansatz for the shape of an infinitely long filament \cite{renardy1994some} as
\begin{subequations}
  \begin{align}
    \h{R}(\h{z},\h{t})&=\h{\tau}^{\alpha_{1}} H_1(\xi_1),\\
    \h{v}(\h{z},\h{t})&=\h{\tau}^{\gamma_{1}} V_1(\xi_1),
  \end{align}
\label{eqn:selfsimi}%
\end{subequations}
where $\h{\tau}=\h{t}_\mr{C}-\h{t}$ is the time to filament breakup. A self-similar variable $\xi_1$ incorporating both the time to filament breakup $\h{\tau}$ and axial position $\h{z}$ is defined as
\begin{equation}
\xi_1\equiv\frac{\h{z}}{\h{\tau}^{\beta_1}}.
\label{eqn:xi}
\end{equation}
The parameters $\alpha_1$, $\beta_1$ and $\gamma_1$ are self-similar exponents. For a Newtonian fluid under a visco-capillary balance, Renardy \cite{renardy1994some} analytically determined that $\beta_1\approx 0.175$, $\alpha_1=1$ and $\gamma_1= \beta_1-1 \approx -0.825$. The self-similar functions $H_1(\xi_1)$ and $V_1(\xi_1)$ as functions of $\xi_1$ can be calculated either analytically or numerically.

We follow a similar procedure to probe the asymptotic solution for the capillarity-driven thinning behavior resulted from the second-order stress component. Specifically, the linear stress component in Eq.~\ref{eqn:sof} is eliminated. In the limit of free inertia and negligible gravitational effects, the momentum equation can be reduced to 
\begin{equation}
\left[\h{R}^2\left( \h{K}+3\mr{Ec}_0 \cdot \h{v}_{,\h{z}}\left|\h{v}_{,\h{z}}\right| \right)\right]_{,\h{z}}=0.
\label{eqn:s2_momentum}
\end{equation}

We presume that the solution to Eq.~\ref{eqn:s2_momentum} has a similar form as Eq.~\ref{eqn:selfsimi} with a different set of parameters $\left(\alpha_2,\gamma_2,\beta_2,H_2,V_2\right)$. As a result, a new self-similar variable is defined as $\xi_2\equiv \h{z}/\h{\tau}^{\beta_2}$. In the remainder of this section, the subscript ``$2$'' in the expressions is dropped for writing simplicity, as we confine our discussion to the second-order stress component. The value of $\alpha$ can be obtained by substituting the ansatz of Eq.~\ref{eqn:selfsimi} into Eq.~\ref{eqn:s2_momentum} as
\begin{equation}
\left\{\h{\tau}^{\alpha}H_{,\xi}+(12\mr{Ec}_0)\h{\tau}^{2\alpha-2}\left[\left(\alpha H-\beta H_{,\xi}\xi\right)\left|\alpha H-\beta H_{,\xi}\xi\right|\right]\right\}_{,\h{z}}=0,
\label{eqn:soF_alpha}
\end{equation}
where the subscript $f_{,\xi}$ and $f_{,\h{z}}$ correspond to the partial derivatives of a function $f$ with respect to $\xi$ and $\h{z}$. This equation is satisfied for any $\h{\tau}$, hence all $\h{\tau}$ terms are canceled out, leading to $\alpha=2$. This component is consistent with the asymptotic solution of Eq.~\ref{eqn:h} with a quadratic dependence on the time to breakup. The calculations of other parameters are as follows. We first define a variable in the Lagrangian frame as
\begin{equation}
\phi\equiv\int_0^{\h{z}} \h{R}^2(\zeta,\h{\tau})\mr{d}\zeta,
\label{eqn:phi}
\end{equation}
which scales with the fluid volume in the range of $0\leq\zeta\leq\h{z}$. As a result, the axial coordinate in the Lagrangian frame becomes a function of $\td{z}\equiv\td{z}(\phi,\h{\tau})$. Using chain rules, we can transform Eq.~\ref{eqn:s2_momentum} to the Lagrangian frame following
\begin{subequations}
  \begin{align}
    \td{z}_{,\phi}&= \td{R}^{-2}(\phi,\h{\tau}),\\
    \td{z}_{,\h{\tau}}&= -\td{v}(\phi,\h{\tau}),\\
    \partial_{\td{z}}&= \td{R}^2(\phi,\h{\tau}) \partial_{\phi},
  \end{align}
\label{eqn:el_transform}%
\end{subequations}
where the tilde symbol indicates functions in the Lagrangian frame in terms of $\phi$ and $\h{\tau}$, \textit{i.e.}, $\tilde{f}(\phi,\h{\tau})\equiv \h{f}[\tilde{z}(\phi,\h{\tau}),\h{\tau}]$. Finally, we rewrite Eq.~\ref{eqn:s2_momentum} in terms of $z(\phi,\h{\tau})$ as
\begin{equation}
-\frac{1}{2}\frac{\td{z}_{,\phi\phi}}{\td{z}_{,\phi}^{3/2}}-3\mr{Ec}_0\left(\frac{\td{z}_{,\h{\tau}\phi}\left|\td{z}_{,\h{\tau}\phi}\right|}{\td{z}_{,\phi}^3}\right)_{,\phi}=0.
\label{eqn:lag_z}
\end{equation}
Using Eq.~\ref{eqn:el_transform} again, we substitute $\td{z}(\phi,\h{\tau})$ in Eq.~\ref{eqn:lag_z} with $\td{R}(\phi,\h{\tau})$ as
\begin{equation}
\td{R}_{,\phi}+12\mr{Ec}_0(\td{R}_{,\h{\tau}}\left|\td{R}_{,\h{\tau}}\right|)_{,\phi}=0.
\label{eqn:lg_sof}
\end{equation}
The mid-plane radius in the Lagrangian frame $\td{h}(\phi, \h{\tau})$ is expected to follow a similar ansatz of the solution as Eq.~\ref{eqn:selfsimi} with a self-similar variable $\chi\equiv \phi/\h{\tau}^{\td{\beta}}$ and a new self-similar exponent $\td{\beta}$. The value of $\td{\beta}$ is determined from the definition of $\phi$ in Eq.~\ref{eqn:phi}. In the leading-order expression, the Lagrangian variable $\phi$ scales with $\phi\sim \h{R}^2 \h{z}\sim \h{\tau}^4H^2(\xi)\h{z}$. Therefore, one can only keep an identical self-similar nature in the Lagrangian frame by letting $\td{\beta}=\beta+4$, which leads to
\begin{subequations}
\begin{align}
\td{h}(\phi,\h{\tau})&=\h{\tau}^{2} \td{H}(\chi),\\
\chi&=\dfrac{\phi}{\h{\tau}^{\beta+4}}.
\end{align}
\label{eqn:selfsimi_Lag}%
\end{subequations}
By substituting Eq.~\ref{eqn:selfsimi_Lag} into Eq.~\ref{eqn:lg_sof} and integrating on both sides, we obtain a differential equation in terms of $\chi$ as
\begin{equation}
\bar{C}(\h{\tau})=\td{H}+12\mr{Ec}_0\left[ 2\td{H}-\left(\beta+4\right)\chi\td{H}_{,\chi} \right]^2.
\label{eqn:sof_c}
\end{equation}
Notice that the right-hand side is a function of $\chi$ only. The absolute sign can be eliminated by presupposing that $\td{h}_{,\h{\tau}}=\h{\tau}\left[2\td{H}-(\beta+4)\chi\td{H}_{,\chi}\right]>0$. This can be rigorously proved by noticing that $\td{h}_{,\h{\tau}}>0$ at $\chi=0$ (as mid-plane radius monotonically decreases in a filament thinning process), and Eq.~\ref{eqn:sof_c} is valid for any $\td{H}(\chi)>0$. We let $\mathcal{K}\equiv(\bar{C}-\td{H})^{1/2}$, and Eq.~\ref{eqn:sof_c} can be rearranged as 
\begin{equation}
\dfrac{\mathcal{K}}{2\sqrt{12\mr{Ec}_0}}=\bar{C}-\mathcal{K}^2+(\beta+4)\chi\mathcal{K}\mathcal{K}_{,\chi}.
\label{eqn:k_int}
\end{equation}
By separating $\mathcal{K}$ and $\chi$, we can integrate on both sides of Eq.~\ref{eqn:k_int} to get
\begin{equation}
\ln\left|\dfrac{\chi}{\chi_0}\right|=(\beta+4)\int_{\mathcal{K}_0}^{\mathcal{K}}\dfrac{\kappa\mr{d}\kappa}{\kappa^2+\dfrac{\kappa}{2\sqrt{12\mr{Ec}_0}}-\bar{C}}.
\label{eqn:k}
\end{equation}
When $\chi_0\rightarrow 0^+$, $\kappa\rightarrow \mathcal{K}_0$, and singularities arise on both sides. To equate Eq.~\ref{eqn:k}, these singularities cancel each other. Therefore, $\mathcal{K}_0$ and $\bar{C}$ can be calculated as
\begin{subequations}
\begin{align}
\mathcal{K}_0&=\dfrac{1}{4\sqrt{12\mr{Ec}_0}(\beta+3)},\\
\bar{C}&=\dfrac{2\beta+7}{192\mr{Ec}_0(\beta+3)^2}.
\end{align}
\label{eqn:k0_c}%
\end{subequations}
From Eq.~\ref{eqn:k} and Eq.~\ref{eqn:k0_c}, we obtain an explicit form of $\chi$ as a function of $\mathcal{K}$ as
\begin{equation}
\begin{split}
\chi(\mathcal{K})=&\left(\dfrac{1}{2\sqrt{12\mr{Ec}_0}}\dfrac{\beta+4}{\beta+3}\right)^{-(\beta+3)/2}\\
&\times \left(\mathcal{K}+\dfrac{1}{4\sqrt{12\mr{Ec}_0}}\dfrac{2\beta+7}{\beta+3}\right)^{(2\beta+7)/2}\left(\mathcal{K}-\mathcal{K}_0\right)^{1/2}.
\end{split}
\label{eqn:xi_L}
\end{equation}
From another perspective, from Eq.~\ref{eqn:el_transform} and Eq.~\ref{eqn:xi_L}, there exists $f(\chi)$ such that $\mathcal{K}^{-2}=f_{,\chi}$. With this notation, we again integrate Eq.~\ref{eqn:k_int} and express $\bar{C}$ in another form as
\begin{equation}
2\sqrt{12\mr{Ec}_0}\bar{C}=\dfrac{\int_{-\infty}^{\infty}\kappa^{-3}\mr{d}\kappa}{\int_{-\infty}^{\infty}\kappa^{-4}\mr{d}\kappa}.
\label{eqn:integralC}
\end{equation}
Finally, we rewrite the integrals in Eq.~\ref{eqn:integralC} in term of hypergeometric integrals \cite{gradshteyn2014table} as
\begin{equation}
    \begin{split}
&\dfrac{(2\beta+7)(-1-\beta)}{2(\beta+3)(-1/2-\beta)}\\
=&\dfrac{F\left[-(5+2\beta)/2,-1-\beta;-1/2-\beta;-7-2\beta\right]}{F\left[-(5+2\beta)/2,-\beta;1/2-\beta;-7-2\beta\right]}.
\end{split}
\label{eqn:hyper}
\end{equation}
Using the bisection method, Eq.~\ref{eqn:hyper} can be numerically solved as
\begin{equation}
\beta=0.212515...,
\label{eqn:beta_2}
\end{equation}
where six significant figures are preserved. The asymptotic evolution of the mid-plane radius $\h{R}_\mr{mid}(\h{\tau})=\h{R}(0,\h{\tau})$ when the second-order stress governs the filament thinning dynamics close to breakup can be instantly obtained as
\begin{equation}
\h{R}_\mr{mid}(\h{\tau})=\dfrac{1}{96\mr{Ec}_0(\beta+3)}\h{\tau}^2=\dfrac{0.00324253...}{\mr{Ec}_0}\h{\tau}^2.
\label{eqn:h0_analytical}
\end{equation}
Therefore, according to Eq.~\ref{eqn:h}, the geometric correction factor $X_\mr{RT}$ can be expressed as
\begin{equation}
X_\mr{RT}=\dfrac{7+2\beta}{4(3+\beta)}=0.577821...
\end{equation}
From Fig.~\ref{fig:diffDe_force}, this constant is qualitatively consistent with the numerical calculation. A more meticulous verification is presented in Sec.~\ref{sec:verify}. 

\subsection{Numerical verification}
\label{sec:verify}
The asymptotic capillary thinning dynamics for the second-order stress analytically obtained in Sec.~\ref{sec:model} are to be rigorously verified by numerical calculations. The simplest idea is to directly substitute Eq.~\ref{eqn:h} into Fig.~\ref{fig:mid_plane_h} to obtain $X_\mr{RT}$. However, the mathematical expression of filament breakup when $\h{R}_\mr{mid}\rightarrow 0$ corresponds to a numerical singularity, thus $\h{t}=\h{t}_\mr{C}$ can never be reached in practice. The determination of $\h{t}_\mr{C}$ requires extrapolation to the limit of $\h{R}_\mr{mid}=0$, and this is done by horizontally shifting the curve of $\h{R}_\mr{mid}(\h{\tau})$ to coincide a power-law trend. Because the asymptotic results of Eq.~\ref{eqn:h} or Eq.~\ref{eqn:h0_analytical} are only manifested close to the filament breakup, or when $\h{\tau}\rightarrow 0$, the extracted power of $\beta_2$ can potentially have large fitting errors, which are more noticeable on a logarithmic scale. To reconcile these shifting errors, we use another dimensionless variable to perform the fitting process in a time-implicit form. For this purpose, we apply the ratio of filament profile curvatures in $z$ and $r$ directions, $\kappa_z$ and $\kappa_r$ evaluated at the mid-plane ($z=0$). Mathematically, this curvature ratio $\Pi$ is defined as
\begin{equation}
\Pi\equiv \frac{\kappa_z}{\kappa_r}=\left|\frac{\h{R}_{,\h{z}\h{z}}}{\left(1+\h{R}_{,\h{z}}^2\right)^{3/2}}\right|/\left|\frac{1}{\h{R}\left(1+\h{R}_{,\h{z}}^2\right)^{1/2}}\right|.
\label{eqn:cr}
\end{equation}
By including the curvatures in both directions, this quantity incorporates the comprehensive information of the filament geometry. Substituting Eq.~\ref{eqn:selfsimi} into Eq.~\ref{eqn:cr}, we can obtain the asymptotic solutions of $\Pi$ with either the linear (with subscript ``N'') or second-order (with subscript ``RT'') stress from Eq.~\ref{eqn:newtonian} and \ref{eqn:h} as
\begin{subequations}
\begin{align}
\Pi_\mr{N}(\h{\tau})&\sim \h{\tau}^{2-2\beta_1}\sim \h{R}_\mr{mid}^{2-2\beta_1(\h{\tau})},\\
\Pi_\mr{RT}(\h{\tau})&\sim \h{\tau}^{4-2\beta_2}\sim \h{R}_\mr{mid}^{2-\beta_2}(\h{\tau}).
\end{align}
\label{eqn:cr_order}%
\end{subequations}
Therefore, by plotting $\Pi(\h{\tau})$ against $\h{R}_\mr{mid}(\h{\tau})$, we can directly obtain an asymptotic solution incorporating the self-similar exponent $\beta$ and avoid manual shifting of the curve.

\begin{figure*}[!ht]
\centering
\includegraphics[width=\textwidth]{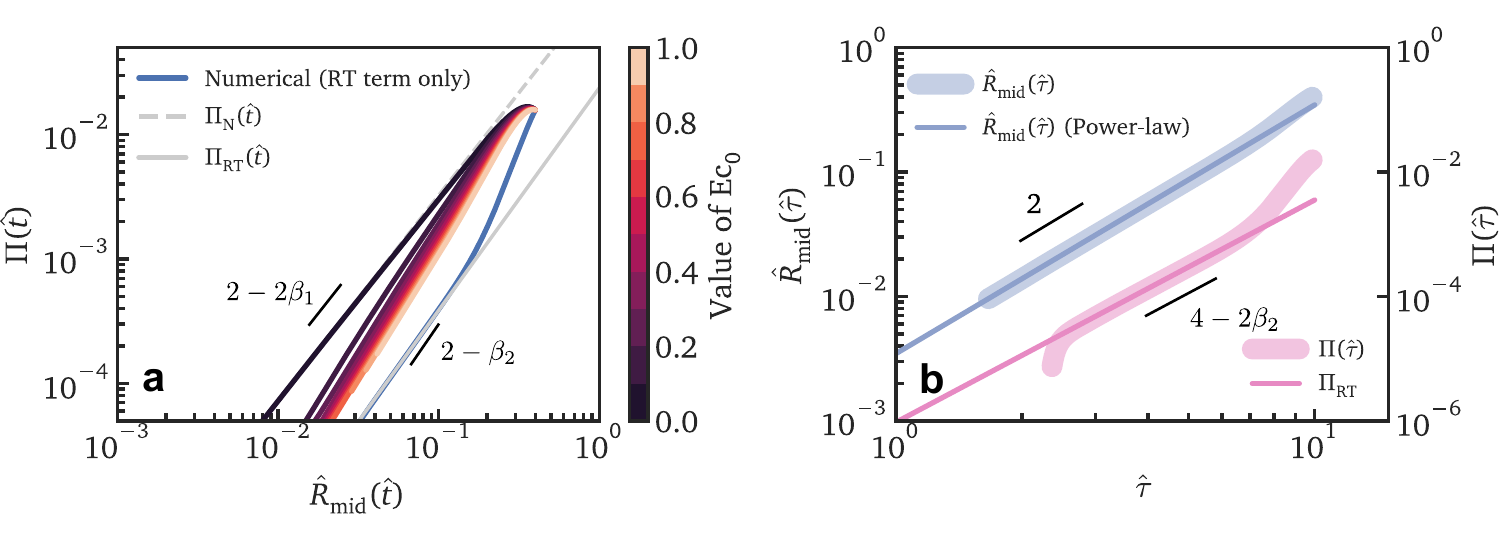}
\caption{(a) Evolution of the curvature ratio $\Pi(\h{t})$ against $\h{R}_\mr{mid}(\h{t})$ at different elasto-capillary numbers. When $\mr{Ec}_0=0$, the numerical calculation for a Newtonian fluid (black line) agrees with the asymptote with a power exponent of $(2-2\beta_1)$ (gray dashed line). The numerical calculation for the second-order stress with vanishing linear stress component (blue line) is also shown to be consistent with the analytical solution of the $(2-\beta_2)$ power-law trend (gray solid line). When $0<\mr{Ec}_0\leq 1$, the evolution of the curvature ratio remains in an intermediate region between the two asymptotic solutions. Geometrically, the filament profile becomes more slender as $\mr{Ec}_0$ increases due to the increasing axial stress contribution from the second-order component. (b) The evolution of mid-plane radius $\h{R}_\mr{mid}(\h{\tau})$ (blue thick line) and curvature ratio at mid-plane $\Pi(\h{\tau})$ (pink thick line) when the capillarity-driven thinning is solely resisted by the second-order stress. The filament breakup time $\h{t}_\mr{C}$ is determined by shifting both curves horizontally such that the curvature ratio $\Pi(\h{\tau})$ coincides with a power trend with a slope of $(4-2\beta_2)$ (pink thin line). The resulting evolution of the mid-plane radius exhibits a quadratic power law against the time to breakup $\h{\tau}$, which coincides with the analytical solution of Eq.~\ref{eqn:h0_analytical} (blue thin line) in both the power exponent and the front factor.}
\label{fig:CRvh}
\end{figure*}

As shown in Fig.~\ref{fig:CRvh}(a), the plot of $\Pi(\h{\tau})$ against $\h{R}_\mr{mid}(\h{\tau})$ follows a power-law trend for the Newtonian fluid ($\mr{Ec}_0=0$, black line) when $\h{R}_\mr{mid}\rightarrow 0$. We also plot the asymptotic solution from Eq.~\ref{eqn:cr_order} with $\beta_1=0.175$ (gray dashed line), and the power-law trend agrees very well with the numerical calculation. As $\mr{Ec}_0$ increases, both the linear and second-order stresses contribute comparably to the capillarity-driven thinning dynamics. The resulting $\Pi(\h{\tau})$ thus deviates from the power law of $(2-2\beta_1)$ for a Newtonian fluid to a smaller value. This decreased curvature ratio geometrically corresponds to a more slender liquid filament profile as shown in Fig.~\ref{fig:h_profile}. We conclude from Fig.~\ref{fig:CRvh}(a) that even a weakly rate-thickening response in the material quantified by a small elasto-capillary number can impact the resulting capillarity-driven thinning dynamics under the balance between capillarity and the second-order stress close to the filament breakup. Meanwhile, the curvature ratio $\Pi(\h{\tau})$ with positive values of $\mr{Ec}_0$ approaches a new power-law asymptote obtained by only retaining the second-order stress (blue line) in solving Eq.~\ref{eqn:euler_force}. This asymptote is well consistent with the analytical result of Eq.~\ref{eqn:cr_order} with a slope of $(2-\beta_2)$ (gray solid line) as $\h{R}_\mr{mid}(\h{\tau})\rightarrow 0$.

Using the results of Fig.~\ref{fig:CRvh}(a), we can obtain the plot of the mid-plane radius $\h{R}_\mr{mid}(\h{\tau})$ against $\h{\tau}$ when the balance between capillarity and the second-order stress dominates the filament thinning dynamics, as shown in Fig.~\ref{fig:CRvh}(b). To obtain a more accurate value of $\h{t}_\mr{C}$ through shifting the curve, we take the evolution of curvature ratio $\Pi(\h{\tau})$ as a reference and plot it in the same figure. We carefully shift both curves horizontally so that the evolution of the curvature ratio coincides with the power-law trend expressed by Eq.~\ref{eqn:cr_order}. As shown in Fig.~\ref{fig:CRvh}(b), when the numerically calculated curvature ratio (pink thick line) overlaps with the desired power-law line (pink thin line) with a power exponent of $(4-2\beta_2)$, the mid-plane radius $\h{R}_\mr{mid}(\h{\tau})$ (blue thick line) exhibits a quadratic power-law trend as $\h{\tau}\rightarrow 0$. Subsequently, we overlap the analytical asymptotic solution of the mid-plane radius close to filament breakup from Eq.~\ref{eqn:h0_analytical} (blue thin line) to the numerical calculation. Comparing both the numerical and analytical results, we obtain excellent agreement in both the quadratic power-law trend and the front factor (within $7\%$ error). Tiny discrepancies between the analytical and numerical calculations are likely to originate from the shifting process, the numerical calculation close to the filament breakup as well as the assumption of an infinitely long filament incorporated in the self-similar ansatz, which is practically unattainable in the numerical calculation. 

\subsection{Calculation of extensional viscosity}

\begin{figure}[!htp]
    \centering
    \includegraphics[width=0.5\textwidth]{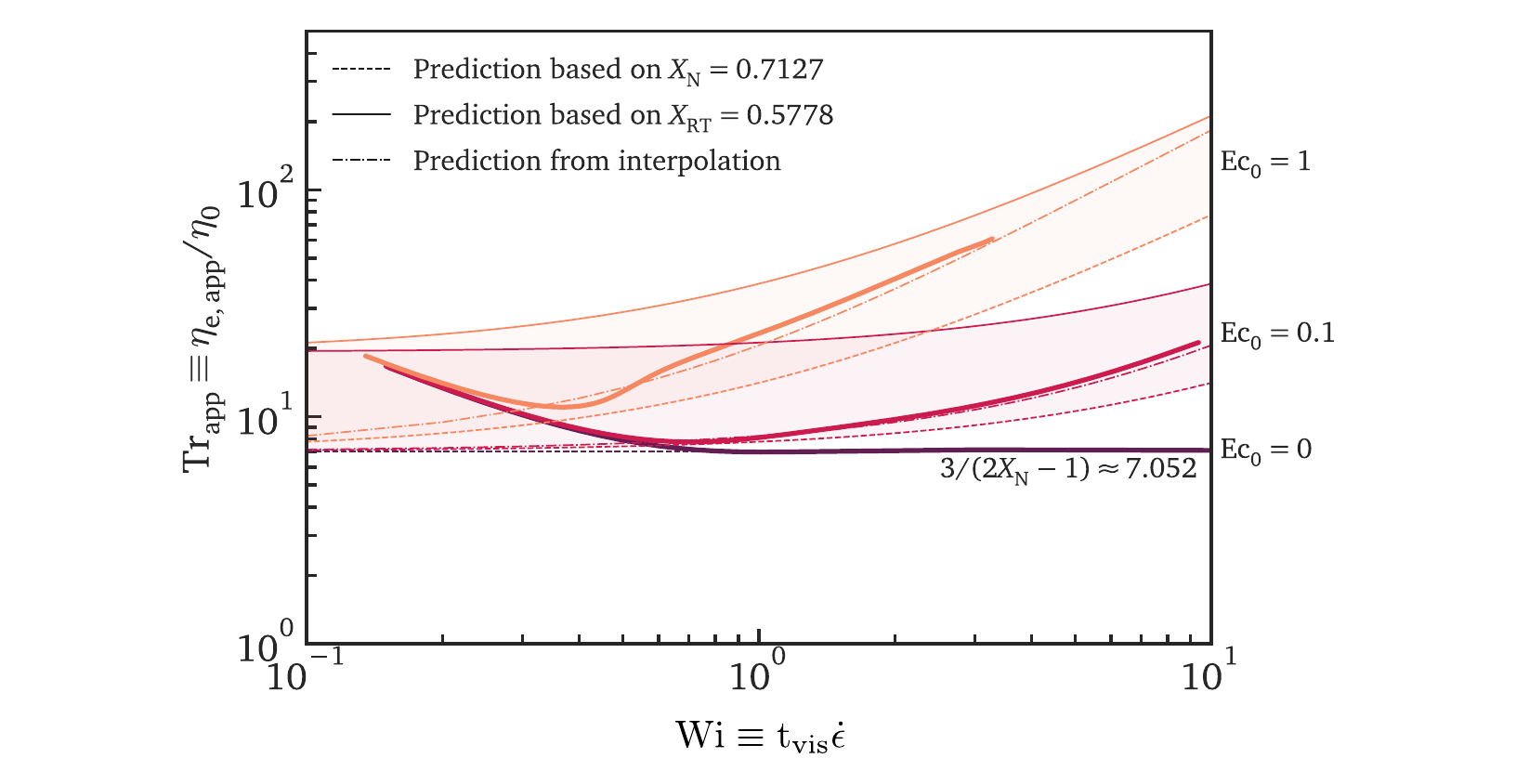}
    \caption{Apparent Trouton Ratio $\mr{Tr}_\mr{app}$ (thick colored lines) against Weissenberg number $\mr{Wi}$. The decreasing apparent viscosity trends at $\mr{Wi}\lesssim 0.5$ are numerical artifacts due to the initial acceleration of the filament thinning. Three elasto-capillary numbers are selected ($\mr{Ec}_0=\numlist{0;0.1;1}$). When $\mr{Ec}_0=0$, the solution for the Newtonian fluid is recovered, and the numerical calculation overlaps with the asymptote of $X_\mr{N}\approx 0.7127$ (violet dashed line).
    When $\mr{Ec}_0>0$, the numerically calculated evolution of $\mr{Tr}_\mr{app}$ is bounded by the two asymptotic solutions of $X=X_\mr{N}\approx 0.7127$ (dashed lines) and $X=X_\mr{RT}\approx 0.5778$ (solid lines). As the Weissenberg number increases, the apparent Trouton ratio $\mr{Tr}_\mr{app}$ progressively deviates from the asymptote of $X=X_\mr{N}$ (thin dashed line) and approaches the asymptote of $X=X_\mr{RT}$ (thin solid lines). This transition occurs approximately at $\mr{Wi}=1/\mr{Ec}_0$. 
    Linearly interpolations of two asymptotic solutions based on the magnitudes of the linear and second-order stresses for each positive $\mr{Ec}_0$ are shown as the dash-dotted lines, and they agree with the numerical calculations.}
    \label{fig:ext_vis_sr}
    \end{figure}

From Eq.~\ref{eqn:lag_force}, the temporal evolution of the geometric correction factor must be accounted for in the extraction of an accurate extensional viscosity from the measured filament thinning profiles for the IRT model. An apparent extensional viscosity $\eta_{\mr{e,app}}$ is calculated from the measured evolution of the mid-plane radius and the surface tension as
\begin{equation}
\eta_\mr{e,app}= \dfrac{\Gamma}{R_\mr{mid}(-2\dot{R}_\mr{mid}/R_\mr{mid})}=\dfrac{-\Gamma}{2\dot{R}_\mr{mid}(\h{t})},
\label{eqn:app_vis}
\end{equation}
where $\dot{\epsilon}=-2\dot{R}_\mr{mid}/R_\mr{mid}$ is the filament strain rate at the mid-plane. The apparent extensional viscosity $\eta_\mr{e,app}$ is a purely experimental measure, hence its magnitude is independent of the evolution of the geometric correction factor $X(t)$. The true extensional viscosity $\eta_\mr{e}$ can be recovered from the apparent extensional viscosities via 
\begin{equation}
\eta_\mr{e}=[2X(t)-1]\eta_\mr{e,app}.
\label{eqn:app_true_connect}
\end{equation}
For a Newtonian fluid, the geometric correction factor is a constant of $X=X_\mr{N}\approx 0.7127$, hence $\eta_\mr{e}\approx 0.4254\eta_\mr{e,app}$. However, for a weakly rate-thickening fluid described by the IRT model, the geometric correction factor varies with time and an accurate true extensional viscosity can be obtained only if the evolution of $X(t)$ is fully characterized. In the IRT model, the dimensionless true extensional viscosity, or the true Trouton ratio can be analytically calculated as $\mr{Tr}(\mr{Wi})\equiv \eta_\mr{e}/\eta_0=3(1+\mr{Ec}_0\cdot\mr{Wi})$, where $\mr{Wi}$ is the dimensionless strain rate, or the Weissenberg number defined in Eq.~\ref{eqn:lag_force}. Give the evolution of $X(t)$ from Fig.~\ref{fig:diffDe_force}, we can plot the dimensionless apparent extensional viscosity, or the apparent Trouton ratio $\mr{Tr}_\mr{app}\equiv \eta_\mr{e,app}/\eta_0$ against $\mr{Wi}$ for different elasto-capillary numbers, as shown in Fig.~\ref{fig:ext_vis_sr}. For plot legibility, we choose three different elasto-capillary numbers, \numlist{0;0.1;1}, corresponding to a Newtonian fluid and two weakly rate-thickening fluids with distinct rates of extensional thickening. Numerical calculations at $\mr{Wi}\lesssim 0.5$ correspond to numerical artifacts due to the initial acceleration of the liquid filament and thus do not reflect the true constitutive responses. 
The evolutions of the apparent Trouton ratio for each $\mr{Ec}_0$ are compared with the two asymptotic solutions with $X=X_\mr{N}\approx 0.7127$ (asymptote with only the linear stress, thin dashed lines) and $X=X_\mr{RT}\approx0.5778$ (asymptote with only the second-order stress, thin solid lines). When $\mr{Ec}_0=0$ (violet), the asymptotic solution for a Newtonian fluid is recovered, and the apparent Trouton ratio from the numerical calculation $\mr{Tr}_\mr{app}=3/(2X_\mr{N}-1)\approx 7.052$ agrees with the asymptotic solution under a visco-capillary balance. When $\mr{Ec}_0>0$, the apparent Trouton ratio $\mr{Tr}_\mr{app}$ is contributed by both the linear and second-order stress components, and hence its magnitude remains between the two asymptotic solutions (shaded area). As the Weissenberg number increases, the apparent Trouton ratio seamlessly evolves from the asymptotic solution of $X=X_\mr{N}$ to $X=X_\mr{RT}$. This transition occurs at a critical Weissenberg number of $\mr{Wi}_\mr{cr}\sim 1/\mr{Ec}_0$ (not plotted).
From Fig.~\ref{fig:ext_vis_sr}, 
at a small elasto-capillary number $\mr{Ec}_0=0.1$ at $\mr{Wi}=10$, the magnitude of $\mr{Tr}_\mr{app}$ from the numerical calculation is $54\%$ larger than the asymptotic solution of $X=X_\mr{N}$. This specific elasto-capillary number and Weissenberg number correspond to a number of automotive lubricants and their typical working conditions \cite{sharma2015rheology,rodd2004capillary}. If we apply the asymptote of $X=X_\mr{N}$ from a Newtonian fluid to recover the true extensional viscosity $\eta_\mr{e}$ using Eq.~\ref{eqn:app_true_connect}, a large error can be generated especially at high strain rates. Previous studies mainly focus on the capillarity-driven thinning dynamics of constitutive models with a single stress component (\textit{e.g.}, a Newtonian fluid \cite{mckinley2000extract}) or in the limit that one particular stress component dominates the filament thinning (\textit{e.g.}, the Oldroyd-B model under an elasto-capillary balance \cite{entov1997effect}). A few studies are available on more complex constitutive models with an interplay of multiple stress components determining the thinning dynamics, but they overwhelmingly neglected the temporal evolution of the geometric correction factor in the extraction of the extensional rheological properties \cite{clasen2006dilute,niedzwiedz2010extensional,wagner2015analytic}.

To render a more accurate measurement of the extensional viscosity for the IRT model, we linearly interpolate the geometric correction factor from the magnitudes of the linear and second-order stresses, which can be expressed as
\begin{equation}
\dfrac{X(\mr{Wi})-X_\mr{N}}{X_\mr{RT}-X_\mr{N}}=\dfrac{\mr{Ec}_0\cdot\mr{Wi}}{1+\mr{Ec}_0\cdot\mr{Wi}}. 
\label{eqn:interpolate_x}
\end{equation}
This expression is applied to predict the apparent extensional viscosities at different $\mr{Ec}_0$, and they are plotted in Fig.~\ref{fig:ext_vis_sr} as dash-dotted lines. From this figure, all interpolations generally agree very well with the numerical calculations (thick colored lines). To recover the true extensional viscosity, Eq.~\ref{eqn:app_true_connect} is applied again with the apparent extensional viscosity as a known quantity obtained from the surface tension and the measured temporal evolution of the mid-plane radius of liquid filament as 
\begin{equation}
\dfrac{\eta_\mr{e}(\dot{\epsilon};\eta_0,k_2)}{2X(\dot{\epsilon};\eta_0,k_2)-1}=\dfrac{3\eta_0+3k_2\dot{\epsilon}}{2X(\dot{\epsilon};\eta_0,k_2)-1}=\eta_\mr{e,app}(\dot{\epsilon}).
\label{eqn:fit_de_interpolation}
\end{equation}
From this equation, the true extensional viscosity can be recovered by numerically fitting the zero-shear viscosity $\eta_0$ and the rate of extensional thickening $k_2$, which are then regrouped into the constitutive equation.

\section{Choice of the best-fit model}
\label{sec:bic}
\subsection{Bayesian information criterion}
Given a measured evolution of the liquid filament profile using the capillary breakup technique, it is necessary to choose the best-fit model to extract a parsimonious set of extensional rheological parameters. We summarize this idea by proposing a robust protocol to choose the best-fit model based on both fitting accuracy and regularization of parameters. We can then extract the accurate extensional rheological properties by fitting the measured data with a prediction line incorporating the constitutive equation and the evolution of the geometric correction factor from the given model.

\begin{figure*}[!htp]
\centering
\includegraphics[width=0.9\textwidth]{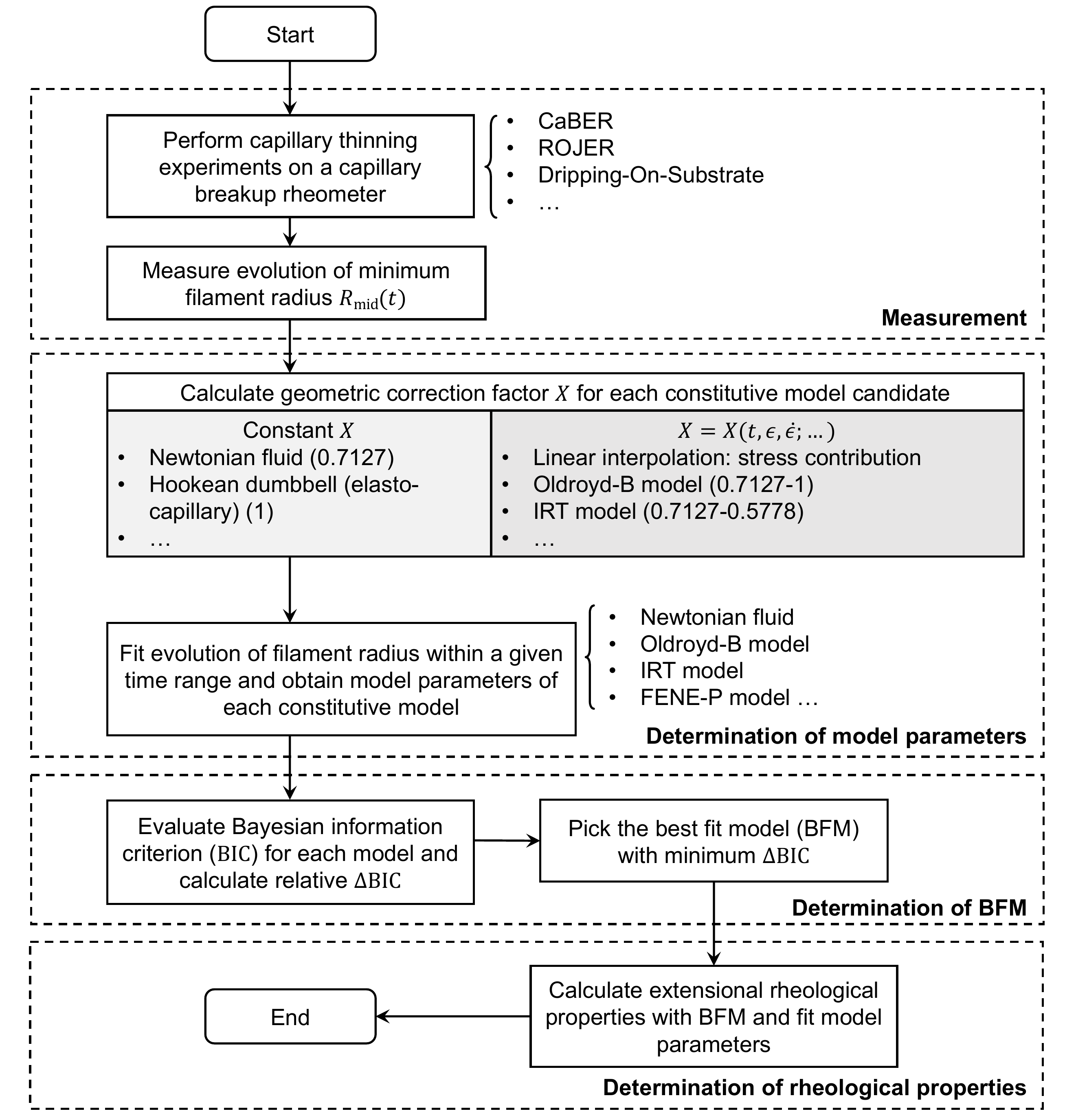}
\caption{Flowchart showing the protocol proposed in this paper to quantify the extensional rheological behavior of a fluid through capillary breakup techniques. Four steps are separately illustrated: (1) The filament thinning of a fluid is initiated on a capillary breakup rheometer, and the resulting evolution of the liquid filament radius or diameter is recorded. (2) The evolutions of the geometric correction factor $X$ for the models to be fitted are characterized using numerical or analytical procedures, and are applied to fit the evolution of filament profile obtained from (1) to obtain the constitutive parameters. (3) The best-fit model (BFM) is determined by calculating the relative Bayesian information criterion ($\Delta$BIC) for each constitutive model candidate and choosing the one with the minimum magnitude. (4) More comprehensive extensional rheological properties are obtained by substituting the fitted constitutive parameters into the best-fit model.}
\label{fig:flowchart}
\end{figure*}

As shown in Fig.~\ref{fig:flowchart}, the protocol is comprised of four steps. First, the capillary thinning of a fluid sample is initiated and the temporal evolution of the resulting liquid filament is measured on an capillary breakup rheometer. 
With an independent measurement of the fluid surface tension, we can fit the evolution of the mid-plane filament radius into a number of constitutive model candidates to extract their model parameters. For each model, the evolution of the geometric correction factor is characterized from numerical or analytical calculations and must be incorporated into the fitting process. The best-fit model is chosen by simultaneously optimizing the fitting accuracy and number of parameters in the constitutive model. To quantify this criterion, we apply the parameter-free Bayesian information criterion (BIC) defined as \cite{schwarz1978estimating}
\begin{equation}
\mr{BIC}=\ln(n)N_{\theta}-2\ln (\h{L}),
\label{eqn:bic}
\end{equation}
where $n$ is the number of data points for fitting, and $N_{\theta}=N_f+1$ is the number of all fitting parameters $\theta=\{p_1,p_2,...,p_{N_f},\sigma^2\}$, which include the model parameters and a variance of $\sigma^2$. To fit a given dataset $\{(x_1,y_1), (x_2,y_2),...,(x_n,y_n)\}$ with a $N_f$-parameter model $f(x;\{p_1,p_2,...,p_{N_f}\})$, we assume that the data at $x=x_i$ satisfies a Gaussian distribution with a mean value of $f(x_i;\{\h{p}_1,\h{p}_2,...,\h{p}_{N_f}\})$ and a variance of $\sigma^2$. We simply need to calculate the best estimator of $\h{\theta}$ such that the likelihood function $\h{L}=p(\{y_i\}|\h{\theta})$ is maximized. The most common form of this likelihood function is taken from the least squares regression (LSR). Mathematically, we can express this likelihood function in a specific form of log-likelihood as
\begin{equation}
    \begin{split}
\ln ({L})=&-\dfrac{n}{2}\ln (2\pi)-\dfrac{n}{2}\ln {\sigma^2} \\
&- \dfrac{1}{2{\sigma}^2}\sum_{i=1}^n \left[y_i-f\left(x_i;\{{p}_1,{p}_2,...,{p}_{N_f}\}\right)\right]^2.
    \end{split}
\label{eqn:mlf}
\end{equation}
The best estimator of parameters $\h{\theta}$ and the maximum likelihood function $\h{L}$ can be obtained by taking the derivatives of Eq.~\ref{eqn:mlf} with respect to each parameter and setting them to vanish. Using LSR, the best estimator of the variance can be expressed as 
\begin{equation}
\h{\sigma}^2=\dfrac{1}{n}\sum_{i=1}^n \left[y_i-f\left(x_i;\{\h{p}_1,\h{p}_2,...,\h{p}_{N_f}\}\right)\right]^2.
\label{eqn:sigma_best}
\end{equation}
Finally, the Bayesian information criterion is calculated from Eq.~\ref{eqn:bic} to~\ref{eqn:mlf} as
\begin{equation}
\mr{BIC}(n,N_F,f)=\ln(n)(N_F+1)+n\left[\ln (2\pi)+1\right]+n\ln(\h{\sigma}^2).
\label{eqn:bic_final}
\end{equation}
Eq.~\ref{eqn:bic_final} thus provides a metric of the ``fitness'' of a constitutive model incorporating both the fitting accuracy (by $\h{\sigma}^2$) as well as the regularization of constitutive parameters. We can thus define the best-fit model to be the one with a minimum value of $\mr{BIC}$.

\subsection{Experiments on complex viscoelastic fluids}
\begin{figure*}[!htp]
    \centering
    \includegraphics{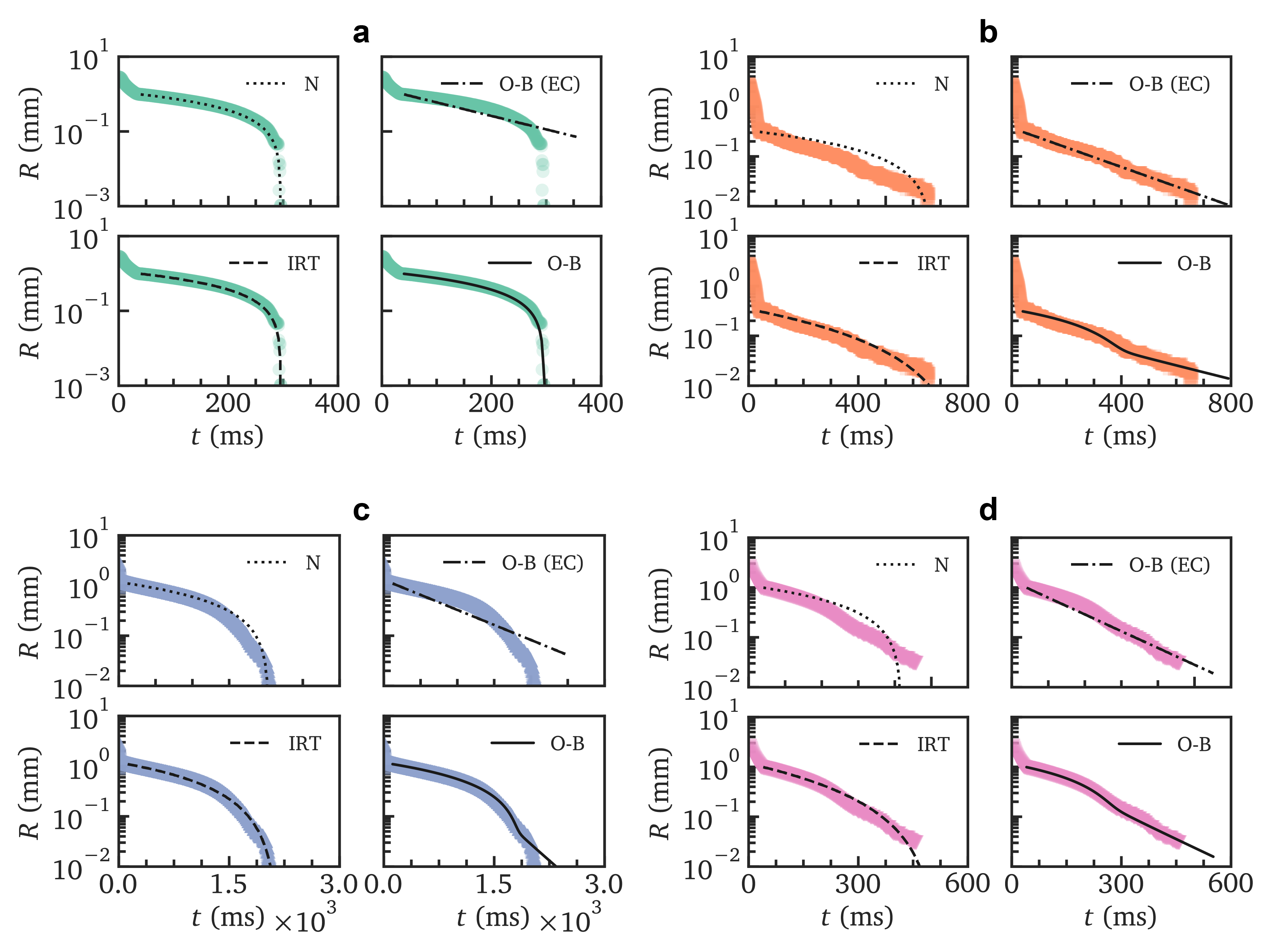}
    \caption{Measured capillarity-driven thinning kinematics for the four selected material systems: (a) Glycerol; (b) PEO/Water (\SI{0.20}{\text{wt\%}}); (c) PIB/hexadecane (\SI{6.47}{\text{wt\%}}); (d) PIB/hexadecane (\SI{4.07}{\text{wt\%}}). In each subplot, the identical experimental data are fitted with four selected constitutive models. N: Newtonian-fluid model (dotted line); O-B (EC): Oldroyd-B model in the elasto-capillary limit (dotted dashed line); IRT: Inelastic Rate-Thickening model (dashed line); O-B: Oldroyd-B model (solid line).}
\label{fig:bic}
\end{figure*}

To demonstrate the validity of this protocol, we measured the filament thinning kinematics of four selected material systems: glycerol, aqueous polyethylene oxide (PEO, $M_\mr{w}\approx \SI{5}{\mega Da}$; Sigma-Aldrich) solutions, and polyisobutylene (PIB, $M_\mr{w}\approx \SI{1}{\mega Da}$, Sigma-Aldrich) solutions dissolved in hexadecane (Sigma-Aldrich). All the selected samples with their concentrations used for demonstration are summarized in Table~\ref{tab:bic_exp}. The measurements are performed on a customized CaBER system with an identical set of geometrical and stretch configuration as in the previous work \cite{duImprovedCapillaryBreakup2021}. The extracted minimal filament radius for each sample is plotted in the subfigures of Fig.~\ref{fig:bic}, and the best-fit model for each sample is determined by the proposed protocol using the experimental data at $t\geq t_\mr{M}=\SI{30}{\ms}$, where $t_\mr{M}$ is the specified motor actuating time. Here, four constitutive equations with varying degrees of viscoelasticity are evaluated to find the BFM for each sample: the Newtonian-fluid model, the Inelastic Rate-Thickening model, the Oldroyd-B model and the Oldroyd-B model in the elasto-capillary limit, denoted as ``N'', ``IRT'', ``O-B'' and ``O-B (EC)'', respectively. According to Eq.~\ref{eqn:app_vis} and \ref{eqn:app_true_connect}, the extensional viscosity is primarily dependent on $\dot{R}(t)$, thus a logarithmic form of the filament radius is used for regression (\textit{i.e.}, $y_i=\ln{R_i}$) for the calculation of $\mr{BIC}$ to generate an unbiased fitting regardless of the magnitude of the filament radius.

\begin{table*}[!htp]
    \centering
    \caption{Selected material systems to demonstrate the proposed statistics-based protocol for the selection of the best-fit constitutive model among the Newtonian fluid (N), the Oldroyd-B model in the elasto-capillary limit [O-B (EC)], the IRT model and the Oldroyd-B (O-B) model. Models with the minimum values of $\mr{BIC}$ are marked in gray.}
    \begin{tabular}{lllllll}
    \toprule
    \multirow{2}{*}{Materials} & $c$ & \multicolumn{4}{c}{$\mr{BIC}$} & \multirow{2}{*}{Figure}\\
    \cmidrule(l r){3-6}
    & ({\text{wt\%}}) & N & O-B (EC) & IRT & O-B \\
    \midrule
    Glycerol & - & \cellcolor{gray!50}\num{-1696.77} & \num{-506.73} & \num{-1658.15} & \num{-1685.81} & \ref{fig:bic}(a) \\ 
    PEO/Water & \num{0.20} & \num{-1464.43} & \cellcolor{gray!50}\num{-5823.78} & \num{-3580.58} & \num{-5104.50} & \ref{fig:bic}(b)  \\ 
    PIB/hexadecane & \num{6.47} & \num{-8695.37} & \num{-5431.68} & \cellcolor{gray!50}\num{-18244.84} & \num{-16578.83} & \ref{fig:bic}(c) \\ 
    PIB/hexadecane & \num{4.07} & \num{-768.64} & \num{-2651.67} & \num{-2325.62} & \cellcolor{gray!50}\num{-4495.58} & \ref{fig:bic}(d) \\ 
    \bottomrule
    \end{tabular}
    \label{tab:bic_exp}
\end{table*}

In each subfigure of Fig.~\ref{fig:bic}, an identical dataset from a particular sample is fitted into the four selected constitutive equations in each quadrant, and the corresponding values of $\mr{BIC}$ are calculated and reported in Table~\ref{tab:bic_exp}, in which the minimum value of $\mr{BIC}$ is highlighted in gray. For the IRT model and the Oldroyd-B model, a time-varying geometric correction factor is applied following the interpolation scheme of Eq.~\ref{eqn:interpolate_x} (see Supplementary Information).

We carefully selected the four material systems such that each of the four model candidates is the BFM for one material. From Fig.~\ref{fig:bic}, it can be demonstrated that the proposed statistics-based criterion provides useful guidance in selecting the BFM with a sufficiently regularized set of parameters. For example, glycerol is well-known to behave as a Newtonian fluid. From Table~\ref{tab:bic_exp}, it is evident that the Newtonian-fluid model outperforms other candidate models in the value of $\mr{BIC}$ due to its fewer number of constitutive parameters, despite that the fitting performance from the IRT model and the Oldroyd-B model appears to the identical, if not better (Fig.~\ref{fig:bic}(a)). This principle of parsimony is also visualized for the aqueous PEO solution with a relatively large molecular weight (Fig.~\ref{fig:bic}(b)), in which the elasto-capillary balance dominates almost the entire lifetime of filament thinning. As a result, the Oldroyd-B model in the elasto-capillary limit can conceivably describe the exponential-thinning trend. In contrast, the capillarity-driven thinning dynamics of the two selected PIB/hexadecane solutions appear to be dominated by multiple stress contributions due to their complex evolution of the minimal filament radius. As the concentration of PIB increases from \SI{4.07}{\text{wt\%}} (Figure~\ref{fig:bic}(d)) to \SI{6.47}{\text{wt\%}} (Figure~\ref{fig:bic}(c)), the best-fit model (BFM), however, switches from the Oldroyd-B model to the more weakly viscoelastic IRT model. This inverse trend of viscoelasticity with the polymer concentration has been identified in the previous work for a number of polymer systems, in which an increasing polymer concentration does not necessarily lead to more pronounced exponential filament thinning (see the Supplemental Information of \cite{duImprovedCapillaryBreakup2021}). Instead, a semi-empirical criterion has been proposed by using a critical crossover elasto-capillary number $\mr{Ec}_\mr{c}=4.7$, below which the material is deemed to be weakly viscoelastic, and the IRT model appears to be more applicable \cite{duImprovedCapillaryBreakup2021}. Here, the crossover elasto-capillary numbers of the two PIB/hexadecane solutions can be calculated as \num{3.6} (\SI{6.47}{\text{wt\%}}; BFM: IRT) and \num{7.8} (\SI{4.07}{\text{wt\%}}; BFM: Oldroyd-B), which are well consistent with the previously proposed criterion. As a result, the statistics-based protocol proposed in this work is sufficiently justified to capture the features in the filament thinning kinematics. More importantly, it is readily applicable to other fitting processes in various rheological characterizations that involve the selection of BFM, and to extract a parsimonious set of fitting parameters more robustly and systematically. 

\section{Conclusion}
In this study, the capillarity-driven thinning dynamics of weakly rate-thickening fluids are placed under scrutiny. This type of fluids has been experimentally characterized on a customized capillary breakup rheometer from a previous study, and their complex rheological response can be appropriately described by an Inelastic Rate-Thickening (IRT) model. This two-parameter model constitutes a Newtonian (linear) stress and an additional non-Newtonian (second-order) stress which scales quadratically with the extensional rate, and both stress components remain comparable in magnitude during the flow deformation. However, an accurate measurement of the constitutive parameters is unattainable from the filament thinning dynamics in the absence of a fully characterized evolution of the geometric correction factor $X$. 

To address this limit, we apply both numerical and analytical procedures to investigate the capillarity-driven thinning dynamics of the IRT model. We apply the numerical scheme proposed by Eggers and Dupont \cite{eggers1994drop}, which is applicable to a slender filament with free inertia. A semi-implicit form of finite difference is used for the time marching. By inspecting the resulting evolutions of the mid-plane filament radius as well as the axial stress components, we find the radius follows a quadratic trend with time to breakup close to the singularity when the second-order stress governs the filament thinning, and the overall geometric correction factor $X$ progressively deviates from the well-studied solution for a Newtonian fluid ($X_\mr{N}\approx 0.7127$) to converge to another constant. This new constant indicates an undiscovered self-similar solution to describe the filament thinning under the balance between capillarity and the second-order stress. To obtain this solution, we follow the analytical procedures of Renardy \cite{renardy1995numerical} for a Newtonian fluid and postulate an identical self-similar ansatz with a different set of parameters. After some mathematical transformation, we obtain a new geometric correction factor of $X_\mr{RT}\approx 0.5778$, which is consistent with the previous numerical calculation. 

The two asymptotic self-similar solutions determined from the first- and second-order stresses are carefully verified by numerical calculations. To quantify the liquid filament profiles at the mid-plane, we calculate the curvature ratio $\Pi\equiv\kappa_z/\kappa_r$ between the axial and radial directions at $z=0$, and plot it against the mid-plane radius as a time-implicit scheme. The two asymptotic power-law trends for the first- and second-order stresses close to breakup agree well with the analytical predictions. Using the curvature ratio as a guidance, we can convert the temporal evolution of the mid-plane radius $\h{R}_\mr{mid}$ from numerical calculations against $\h{\tau}$, the time to filament breakup, and both the numerical and analytical results overlap with each other reasonably well.
From the newly obtained self-similar solutions, we can obtain the apparent extensional viscosity $\eta_\mr{e,app}$ as a practical measure of the extensional viscosity from capillary breakup techniques. To recover the true extensional viscosity, we linearly interpolate the geometric correction factor $X$ based on the relative magnitude of the first- and second-order stresses. This dynamical geometric correction factor is then incorporated into the fitting process to produce a more accurate measurement of the extensional rheological properties.


Finally, we propose a robust measuring protocol to efficiently and accurately quantify the extensional rheology of an unknown fluid using capillarity-driven breakup techniques for a general constitutive model. In this protocol, the measured filament thinning profiles are fitted with a selected range of constitutive model candidates, and the evolution of the geometric correction factor for each model is incorporated in the fitting process. The Bayesian information criterion (BIC) is then applied to evaluate the goodness of fitting for each model, from which the best-fit model is determined with an optimized fitting accuracy and a set of well-regularized parameters. The extensional rheological properties of the measured fluid can be obtained by substituting the fitted constitutive parameters into the best-fit model. We apply this protocol to a number of complex viscoelastic fluids. The fitting results and the selection of the best-fit model are well consistent with the previous non-statistics-based criteria. Furthermore, this protocol can be readily automated with computer-aided imaging and data processing, and can find potential applications in many industrial processes, in which the extensional rheology for a large number of fluids can be characterized more efficiently and accurately.

\section*{Acknowledgment}
J.D. and G.H.M. would like to thank Ford Motor Company for the financial support on this project. J.D. would like to thank Prof. Christian Clasen from KU Leuven for inspiring discussions and suggestions on the results.

\bibliography{secondOrderCorrection}

\end{document}